\documentclass{aa}

\usepackage{graphicx}
\usepackage{txfonts}
\usepackage{lipsum}
\usepackage{subcaption}         % necessary for continued figures, example in section 3
\usepackage{lscape}             % to rotate a single page table, example in appendix.
\usepackage{placeins}           % useful with \FloatBarrier, to keep 
\usepackage{xcolor}

\def\chisqr{\hbox{$\chi^2_{\rm r}$}}
\def\msun{\hbox{${\rm M}_{\odot}$}}

\def\rsun{\hbox{${\rm R}_{\odot}$}}

\def\mstar{\hbox{$M_{\star}$}}
\def\rstar{\hbox{$R_{\star}$}}

\def\teff{\hbox{$T_{\rm eff}$}}
\def\logg{\hbox{$\log g$}}

\def\vD{\hbox{$v_{\rm D}$}}

\def\kms{\hbox{km\,s$^{-1}$}}

\def\vsini{\hbox{$v \sin i$}}

\def\mic{\hbox{$\mu$m}}

\def\emr{}
\def\Bl{\hbox{$B_{\rm \ell}$}}
\def\Bd{\hbox{$B_{\rm d}$}}

\def\degr{\hbox{$^\circ$}}

\def\Prot{\hbox{$P_{\rm rot}$}}

\begin{document}

%%%%%%%%%%%%%%%%%%% TITLE PAGE %%%%%%%%%%%%%%%%%%%

\title{Full Stokes magnetometry of the active M dwarfs AU~Mic and EV~Lac with SPIRou} 

   \author{J.-F.~Donati\inst{1}
      \and P.I.~Cristofari\inst{2}
      \and B.~Klein\inst{3}
      \and B.~Finociety\inst{4}
      \and C.~Moutou\inst{1}
          }
   \institute{Univ.\ de Toulouse, CNRS, IRAP, 14 avenue Belin, 31400 Toulouse, France\\ \email{jean-francois.donati@irap.omp.eu}
         \and Leiden Observatory, Leiden University, Niels Bohrweg 2, 2333 CA Leiden, the Netherlands
         \and Department of Physics, University of Oxford, Oxford OX13RH, UK 
         \and ACRI-ST, 260 Route du Pin Montard, BP 234, 06904 Sophia-Antipolis
             }

% These dates will be filled out by the publisher
\date{Submitted 2025 May 07 -- Accepted 2025 June 17} 

% Abstract of the paper
\abstract{
We report in this paper circularly and linearly polarized observations of the young active M dwarfs AU~Mic and EV~Lac with the near-infrared SPIRou 
spectropolarimeter at the Canada-France-Hawaii Telescope, collected from August to October 2023 over a few rotation cycles of both stars.  Applying Least-Squares 
Deconvolution (LSD) to our spectra, we detected Zeeman signatures in circular (Stokes $V$) and linear (Stokes $QU$) polarization, and Zeeman broadening 
in unpolarized (Stokes $I$) LSD profiles, all exhibiting clear rotational modulation.  Using the stellar surface tomographic technique of Zeeman-Doppler imaging on 
our sets of observations, along with a simple parametric description of how the small-scale and large-scale fields relate to each other, we recovered the magnetic 
topologies of AU~Mic and EV~Lac successively from LSD Stokes $V$, Stokes $IV$ and Stokes $IVQU$ profiles, to investigate how the reconstructed maps evolve as we 
provide more information, and ultimately infer reliable magnetic maps of both stars.  We find that AU~Mic hosts a fairly simple and mostly poloidal 
large-scale field aligned with the rotation axis within about 10\degr, whereas that of EV~Lac is more complex, stronger and less axisymmetric.  
Both stars feature intense small-scale fields, of about 4~kG for AU~Mic and 6~kG for EV~Lac when averaged over the whole stellar surface.  Stokes $QU$ Zeeman 
signatures allow one to reconstruct stellar magnetic fields more reliably, and are especially useful for stars with more complex fields and low \vsini\ like EV~Lac. 
} 

\keywords{stars: magnetic fields -- stars: imaging -- stars: low-mass -- stars: individual: AU~Mic, EV~Lac  -- techniques: polarimetric} 

\maketitle

%%%%%%%%%%%%%%%%%%%%%%%%%%%%%%%%%%%%%%%%%%%%%%%%%%

%%%%%%%%%%%%%%%%% BODY OF PAPER %%%%%%%%%%%%%%%%%%

\section{Introduction}
\label{sec:int}

Magnetic fields play a significant role in the life of low-mass stars, especially during their formation stage when the newly born stars build up from huge 
molecular clouds of interstellar material and magnetically interact with their circumstellar environment, then throughout their main sequence lifetimes, e.g., 
when cool stars loose most of their angular momentum through magnetized winds \citep{Donati09,Reiners12,Kochukhov21}.  These magnetic fields are produced 
through dynamo processes, i.e., amplified from a small seed field by the interplay of rotation and cyclonic convection operating in stellar convective envelopes 
for partly-convective G, K and early M stars, and throughout their interiors for fully-convective M dwarfs.  The Rossby number, i.e., the ratio of the stellar 
rotation rate to the convective turnover time, is a key parameter for dynamo processes, with magnetic field strength scaling with inverse Rossby number 
until it saturates for stars with Rossby numbers below about 0.1 \citep[e.g.,][]{Reiners12,Vidotto14b,See15,Reiners22}.  

Magnetic fields of M dwarfs are particularly interesting for studying dynamo processes in low-mass stars other than the Sun given the large range of their 
rotation rates, spanning from under 1~d to over 100~d \citep[e.g.,][]{Morin10,Donati23b}, and the wide variety in their internal structures, ranging from partly 
convective for mature early-M dwarfs to fully convective to late-M or pre-main-sequence dwarfs.  The recent discovery that very slowly rotating late-M dwarfs can 
host strong large-scale fields \citep{Lehmann24} is quite surprising in this respect.  Magnetic fields measurements from spectropolarimetric observations have 
also shown to be particularly efficient at diagnosing rotational modulation of low-mass stars \citep{Hebrard16,Fouque23,Donati23b}, including for the Sun itself 
\citep{Rescigno24}.  They are also quite useful to investigate magnetic interactions with potential close-in planets \citep[e.g.,][]{Strugarek15,Kavanagh21}, and to 
study the impact of large-scale magnetic fields of planet-hosting stars on the habitability of their close-in planets \citep{Vidotto13}.  

Stellar magnetic fields are usually characterized using the Zeeman effect, i.e., the broadening of unpolarized lines and the circular (Stokes $V$) and linear 
(Stokes $QU$) polarization signatures in line profiles that magnetic fields induce in stellar spectra.  In low-mass stars other than the Sun, this method was first 
used to detect the small-scale tangled fields from unpolarized (Stokes $I$) spectra \citep[e.g.,][]{Robinson80,Saar85}, then the large-scale 
fields through Stokes $V$ spectra \citep[e.g.,][]{Donati92a,Donati97b}.  Tomographic techniques inspired from medical imaging 
\citep{Skilling84} were then applied to observations of low-mass stars to infer their large-scale magnetic fields from time-series of rotationally modulated 
polarized Zeeman signatures \citep[e.g.,][]{Donati92,Donati06b,Morin08b,Morin10,Donati23,Lehmann24,Bellotti24} and sometimes their small-scale magnetic 
fields \citep[e.g.,][]{Kochukhov23}.  Such studies demonstrated that dynamos of low-mass stars share similarities with the one operating in 
the Sun, e.g., with large-scale magnetic topologies exhibiting global polarity switches \citep[e.g.,][]{Fares09,BoroSaikia18,Lehmann24}, but also significant 
differences, e.g., with the toroidal component of the magnetic field being often straightforwardly detected at the surface of the star \citep{Donati92, Donati09}.  

Among all magnetic M dwarfs, AU~Mic and EV~Lac are prototypical examples of key interest for magnetic studies of low-mass stars.  With an age of only about 20~Myr, 
AU~Mic is a pre-main-sequence (PMS) M1 dwarf known to host a debris disk, that was recently shown to also harbour a multi-planet system with at least two transiting 
planets \citep[e.g.,][]{Boccaletti15,Martioli21}.  Its activity and magnetic field are quite strong, as expected from its young age and relatively short rotation 
period \citep{Kochukhov20,Cristofari23,Donati23}.  With an even shorter rotation period, EV~Lac is another extremely active M4 dwarf long known for its exceptional 
rate of energetic flaring, and for its strong small-scale and apparently atypical large-scale magnetic fields \citep[e.g.,][]{Johns96,Favata00,Reiners07,Morin08b}.  

The most recent 
magnetic analysis of AU~Mic showed that modeling the large-scale field from Stokes $V$ data alone can lead to oversimplified magnetic topologies, likely missing 
not only small-scale structures but potentially large-scale features as well \citep{Donati23}.  
{\emr The importance of including Stokes $QU$ Zeeman signatures whenever possible to ensure a more reliable reconstruction of large-scale stellar magnetic topologies 
had been emphasized more than two decades ago in the particular case of chemically peculiar stars \citep{Kochukhov04}, then for solar-type stars about 10~years ago 
\citep{Rosen15}, but the relative faintness of these signatures (versus Stokes $V$ ones) renders them difficult to detect and thus to exploit in practice.  SPIRou observations 
in the nIR where the Zeeman effect is stronger should help a lot in this respect.} For these reasons we initiated a new programme focussing on AU~Mic 
and EV~Lac and based on near-infrared (nIR) spectropolarimetric observations recorded in all four Stokes parameters with SPIRou at the Canada-France-Hawaii Telescope 
(CFHT) atop Maunakea \citep{Donati20}, in order to obtain a more precise description of their magnetic topologies.  

After recalling the main properties of AU~Mic and EV~Lac in Sec.~\ref{sec:par} and describing our new spectropolarimetric SPIRou observations in 
Sec.~\ref{sec:obs}, we outline our modeling of the magnetic topologies of both stars using the stellar tomographic technique of Zeeman-Doppler imaging (ZDI) in 
Sec.~\ref{sec:zdi}.  We summarize our main results and discuss their implications in Sec.~\ref{sec:dis}.

\section{The M dwarfs AU~Mic and EV~Lac}
\label{sec:par}

AU~Mic (Gl~803) is a $\simeq$20~Myr PMS M1 dwarf known for its prominent debris disc with fast-moving features \citep{Boccaletti15,Boccaletti18,Gallenne22,Lawson23}, 
its strong magnetic field \citep[e.g.,][]{Kochukhov20,Klein21,Reiners22,Cristofari23,Donati23,Donati25}, intense activity level \citep[with starspots, flares and 
a possible activity cycle, e.g.,][]{Ibanez19,Cale21,Feinstein22,Bloot24,Waalkes24}, and a multiple planet system featuring at least two transiting Neptune-like planets and 
another two candidate planets \citep[][]{Plavchan20,Klein21,Cale21,Martioli21, Zicher22,Klein22,Szabo22,Donati23,Wittrock23,Mallorquin24,Boldog25,Yu25}.  
With a mass and radius respectively equal to $\mstar=0.60\pm0.04$~\msun\ and $\rstar=0.82\pm0.02$~\rsun, AU~Mic is still contracting towards the main sequence and 
has likely started to develop an inner radiative core \citep{Donati23}.  Seen almost equator-on, AU~Mic rotates in $\Prot=4.86$~d, with its spectral lines exhibiting 
significant rotational broadening \citep[$\vsini=8.5\pm0.2$~\kms,][]{Donati23}.  

\begin{table*}[ht!] 
\caption{Parameters of AU~Mic and EV~Lac used in our study} 
\centering 
\resizebox{\linewidth}{!}{  
\begin{tabular}{ccccccc}
\hline
                     && \multicolumn{2}{l}{AU~Mic}             && \multicolumn{2}{l}{EV~Lac} \\
\hline
distance (pc)        && $9.714\pm0.002$ & \citet{Gaia21}       && $5.0515\pm0.0005$ & \citet{Gaia21}   \\
\teff\ (K)           && $3665\pm31$     & \citet{Cristofari23} && $3340\pm31$   & \citet{Cristofari23} \\
\mstar\ (\msun)      && $0.60\pm0.04$   & \citet{Donati23}     && $0.32\pm0.02$ & \citet{Cristofari23} \\
\rstar\ (\rsun)      && $0.82\pm0.02$   & \citet{Donati23}     && $0.31\pm0.02$ & \citet{Cristofari23} \\
\Prot\ (d)           && 4.86            & \citet{Donati23}     && 4.37          & \citet{Morin08b}     \\ 
\vsini\ (\kms)       && $8.5\pm0.2$     & \citet{Donati23}     && $3.1\pm0.4$   & from \Prot, \rstar\ and $i$ \\ 
$i$ (\degr)          && 80              & assumed for ZDI      && 60            & assumed for ZDI      \\ 
<$B_s$> (kG)         && $2.61\pm0.05$   & \citet{Cristofari23} && $4.52\pm0.07$ & \citet{Cristofari23} \\
Rossby number $Ro$   && 0.14            & \citet{Wright18}     && 0.058         & \citet{Wright18}     \\ 
\hline
\end{tabular}}
\label{tab:par}
\end{table*}

EV~Lac (Gl~873) is an M4 dwarf extensively studied over the last decades for its exceptional flaring properties, known to exhibit some of the most energetic flares 
ever reported for a main-sequence star \citep{Favata00,Paudel24}, making it an ideal laboratory to investigate the extreme magnetic activity that M dwarfs are able to 
trigger and its potential impact on the habitability of nearby planets and on the emergence of life \citep{Feinstein22,Paudel24}.  With a rotation period of 
$\Prot=4.37$~d \citep[e.g.,][]{Morin08b}, shorter than that of AU~Mic, EV~Lac is younger than 1~Gyr \citep{Engle24}, with a mass and radius respectively equal to 
$\mstar=0.32\pm0.02$~\msun\ and $\rstar=0.31\pm0.02$~\rsun\ \citep{Cristofari23}.  EV~Lac hosts a strong small-scale magnetic field of several kG 
\citep{Johns96,Reiners07,Reiners22,Cristofari23} and an atypical large-scale field with a dipole component reported to be strongly tilted with respect to the stellar 
rotation axis \citep{Morin08b, Bellotti24}.  According to evolutionary models \citep{Baraffe15,Feiden16}, EV~Lac is located 
very close to the threshold between partly and fully convective M dwarfs;  depending on its exact mass and radius, EV~Lac may have just started developing a small 
radiative core if no longer fully convective.  This makes EV~Lac and AU~Mic structurally similar, despite their having different masses and radii.  

The large-scale and small-scale magnetic field of AU~Mic and EV~Lac have been studied a lot, including from SPIRou observations collected over 
several seasons \citep[][]{Klein21,Cristofari23,Donati23,Bellotti24,Cristofari25,Donati25}.  Both the longitudinal field \Bl\ (i.e., the line-of-sight projected 
component of the large-scale field averaged over the visible stellar hemisphere, measured from Stokes $V$ line profiles) and the small-scale field $B_s$ (estimated from the broadening 
of Stokes $I$ line profiles) exhibit rotational modulation.  However, these studies either exploited Stokes $V$ or Stokes $I$ data, sometimes both, 
but never observations in all four Stokes parameters.  The magnetic dynamos of AU~Mic and EV~Lac presumably operate in a saturated regime given their Rossby 
numbers $Ro$ respectively equal to 0.14 and 0.058 \citep{Wright18}.  The potential interaction of AU~Mic's large-scale magnetic field on its transiting planets has 
also been studied in several recent papers \citep{Kavanagh21,Alvarado22,Klein22}.  

The main parameters of both stars are recalled in Table~\ref{tab:par}.

\section{SPIRou observations}
\label{sec:obs}

For this study, we observed AU~Mic between 2023 August~02 and September~01, and EV~Lac between 2023 September 26 and October 06, with the SPIRou nIR spectropolarimeter 
\citep{Donati20} at CFHT, within a dedicated programme (RUNID \#23BF08, PI B.~Finociety) complemented by additional exposures on AU~Mic recorded within the SPICE Large Program 
(RUNID \#23BP45, PI J.-F.~Donati).  SPIRou collects unpolarized and polarized stellar spectra, covering a wavelength interval of 0.95--2.50~\mic\ at a resolving power 
of 70\,000 in a single exposure.  With our main programme, we obtained Stokes $IQU$ spectra of AU~Mic and $IQUV$ observations of EV~Lac, whereas Stokes $IV$ exposures 
of AU~Mic were secured within SPICE as part of our long-term monitoring effort on this key target \citep[][]{Donati23,Donati25}.  
Polarization observations with SPIRou usually consist of sequences of four sub-exposures, with each sub-exposure corresponding to different azimuths of the Fresnel rhomb 
retarders of the SPIRou polarimetric unit.  This procedure was shown to succeed in removing systematics in polarization spectra \citep[to first order, see, e.g.,][]{Donati97b}.  
Each recorded sequence yields one Stokes $I$ and one polarized spectrum (either $V$, $Q$ or $U$ depending on the selected Stokes parameter), as well as one null 
polarization check (called $N$) used to diagnose potential instrumental or data reduction issues.  
We recorded a total of 40 polarization sequences for AU~Mic (16 for Stokes $V$ and 12 for each of Stokes $Q$ and $U$) and 27 exposures for EV~Lac (9 for each Stokes $V$, $Q$ 
and $U$ parameter), with typically a complete set of Stokes parameters collected in most clear nights.  The full log of our observations is provided in Tables~\ref{tab:log-au} 
and \ref{tab:log-ev} in Appendix~\ref{sec:appA}.   

All SPIRou spectra of AU~Mic and EV~Lac were processed with \texttt{Libre ESpRIT}, the nominal reduction pipeline of ESPaDOnS at CFHT, optimized for spectropolarimetry 
and adapted for SPIRou \citep{Donati20}.  Subsequently, we applied Least-Squares Deconvolution \citep[LSD,][]{Donati97b} to the reduced spectra, with a line mask computed 
with the VALD-3 database \citep{Ryabchikova15} for a set of atmospheric parameters roughly matching those of AU~Mic and EV~Lac ($\teff=3750$~K and 3500~K respectively, 
both with $\logg=5.0$).  As in previous studies, we only selected atomic lines deeper than 10 percent of the continuum level, for a total of $\simeq$1500 lines of average 
wavelength and Land\'e factor equal to 1750~nm and 1.2.  Our SPIRou data were also processed with \texttt{APERO}, the nominal SPIRou reduction pipeline \citep{Cook22}, 
then analyzed with the line-by-line technique \citep[LBL,][]{Artigau22}, yielding, for both stars and for each observing night, the differential temperatures $dT$ 
estimated from the variation of spectral lines with respect to their median profile \citep{Artigau24}.   We used our $dT$ measurements to infer a relative photometric light 
curve at SPIRou wavelengths (with temperature changes converted into brightness fluctuations with the Planck function), to be adjusted with ZDI along with the LSD profiles of 
both stars (see Sec.~\ref{sec:zdi}).  We obtain light curves at SPIRou wavelengths with full amplitudes of 2.2 and 1.3 percent for AU~Mic and EV~Lac 
respectively during our observations, about twice smaller than those measured with TESS in the $I$ band at other epochs.  

To compute the LSD Stokes $I$ and $V$ profiles, we adopted the usual line weights \citep[respectively 
equal to $d$ and $g\lambda d$ where $g$, $\lambda$ and $d$ denote the Land\'e factor, wavelength and relative depth of the line in the mask, see][]{Donati97b}, and the 
better suited weight $g^2 \lambda^2 d$ to derive LSD Stokes $Q$ and $U$ profiles \citep[following the logic of][]{Donati97b}.  
The noise levels $\sigma_P$ in the resulting Stokes LSD profiles range from 0.58 to 0.95 for AU~Mic (median 0.70, in units of $10^{-4} I_c$ where $I_c$ denotes the continuum 
intensity);  for EV~Lac, $\sigma_P$ ranges from 0.62 to 0.90 (median 0.70) for Stokes $Q$ and $U$ LSD profiles, and from 1.88 to 2.53 (median 2.0) for Stokes $V$ profiles.  

LSD Stokes $V$ signatures are unsurprisingly detected in both stars, as in previously published SPIRou observations \citep{Donati23,Bellotti24}.  The Zeeman broadening of 
LSD Stokes $I$ profiles is particularly obvious, as already discussed in the particular case of AU~Mic \citep{Donati23}.  LSD Stokes $Q$ and $U$ signatures are unambiguously 
detected for both stars in most observations (see Sec.~\ref{sec:zdi}), which confirms the feasibility of full Stokes magnetometry with SPIRou, at least for strongly magnetic 
low-mass stars that are bright enough to yield the requested signal-to-noise ratios (SNRs).  Phases and rotation cycles are derived assuming a rotation period of 
$\Prot=4.86$~d and 4.37~d for AU~Mic and EV~Lac respectively (see Table~\ref{tab:par}), counting from an arbitrary starting BJD of 2\,459\,000 \citep[as in][for AU~Mic]{Donati23}.  

We finally estimated the Zeeman broadening of atomic and molecular lines with ZeeTurbo \citep[as in][]{Cristofari23,Cristofari25} from the individual Stokes $I$ 
spectra of both stars, and derived for each observing night the corresponding values of the average small-scale magnetic field over the visible stellar disc <$B_s$>, listed in 
Tables~\ref{tab:log-au} and \ref{tab:log-ev} along with the inferred values of $dT$.

\section{Zeeman-Doppler Imaging}
\label{sec:zdi}

The goal is now to reconstruct the magnetic topologies of AU~Mic and EV~Lac from the sets of rotationally modulated LSD Stokes profiles derived in the previous section, using 
ZDI.  There are several ways to tackle this task, as previously outlined in the particular case of AU~Mic \citep{Donati23}.  The simplest way is to fit LSD Stokes $V$ profiles 
only, making sure that the average model Stokes $I$ profile is consistent with observations (in both width and equivalent width).  This is the classical method used in most 
ZDI studies, best suited to stars with weak to moderate magnetic fields for which LSD Stokes $I$ profiles exhibit no clear magnetic broadening \citep[e.g.,][]{Lehmann24}, but 
less so for those with stronger fields and clear Zeeman broadening in LSD Stokes $I$ profiles \citep[as discussed in][]{Donati23}.  The second option is to fit both Stokes $I$ 
and Stokes $V$ profiles with ZDI, thereby consistently modeling the Zeeman broadening and circular polarization signatures of line profiles \citep[as achieved for AU~Mic 
in][]{Donati23};  in this case one can also reconstruct brightness inhomogeneities at the surface of the star, from LSD Stokes $I$ profiles (and photometric data when available).  
The third and most complete approach is to use all Stokes $IVQU$ data along with photometry to derive the most accurate description of the parent magnetic topology and surface 
brightness distribution with ZDI;  this method is however not easy given the small amplitudes of the Stokes $QU$ Zeeman signatures in most low-mass stars except the strongly 
magnetic ones, and was only implemented so far for a single low-mass star other than the Sun \citep{Rosen15}.  

We describe in the following paragraphs what we obtain for both AU~Mic and EV~Lac with these three different approaches using 
our SPIRou LSD profiles, complemented with the light curves derived from $dT$ when fitting LSD Stokes $I$ profiles (in the second and third methods).

\subsection{Method description}
\label{sec:met}

In practice, ZDI operates as described in previous papers based on SPIRou data, starting from a small seed magnetic field and a blank brightness distribution, proceeding 
iteratively by adding information on the image as it explores the parameter space using conjugate gradient techniques.  At each iteration, ZDI compares the synthetic Stokes profiles 
of the current magnetic image with observed ones, and loops until it reaches the requested level of agreement with the data (i.e., a given reduced chi-square \chisqr).  
The surface of the star is typically decomposed into 5000 grid cells, each associated with a value 
of the local surface brightness relative to that of the quiet photosphere.  The magnetic topology is described through a spherical harmonics (SH) expansion using the formalism of \citet{Donati06b} 
in its revised implementation \citep{Lehmann22, Finociety22}, where the poloidal and toroidal components of the vector field are expressed with three sets of complex SH coefficients, 
$\alpha_{\ell,m}$ and $\beta_{\ell,m}$ for the poloidal component, and $\gamma_{\ell,m}$ for the toroidal component (with $\ell$ and $m$ denoting the degree and order of the corresponding 
SH term in the expansion).  Given the modest rotational broadening \vsini\ of both stars (see Table~\ref{tab:par}), we limit the SH expansion describing the field to $\ell=10$, 
i.e., larger than the usual $\ell=5$ option to potentially account for the smaller magnetic features that full Stokes magnetometry can give access to \citep{Rosen15}.  

{\emr With about four spectrally resolved elements (of size $\simeq$4~\kms) throughout line profiles for AU~Mic and two for EV~Lac, we only have $\simeq$160 independent Stokes $VQU$ data 
points for AU~Mic and $\simeq$54 for EV~Lac, to reconstruct the 180 SH complex coefficients, i.e., the 360 parameters, describing the large-scale field.  
It implies that the inversion problem is ill-posed and that some regularization is needed, especially in the case of EV~Lac, even for reconstructions from Stokes $IVQU$ profiles supposed to 
have a unique solution in ideal conditions \citep{Piskunov05}.  ZDI chooses the simplest among the multiple solutions, i.e., the one with minimum information or maximum entropy that 
matches the data at the requested \chisqr\ level, following the approach of \citet[][]{Skilling84}.} 

To compute local synthetic Stokes $IVQU$ profiles from each grid cell, we use again Unno-Rachkovsky's equation of the polarized radiative transfer equation in a plane-parallel 
Milne-Eddington atmosphere \citep{Landi04}.  We then integrate the spectral contributions from all visible grid cells (assuming a linear center-to-limb darkening law for the continuum, 
with a coefficient of 0.3) to obtain the global synthetic profiles at each observed rotation phase.  The mean wavelength and Land\'e factor of our synthetic profiles are mirrored 
from those of our LSD profiles, i.e., 1700~nm and 1.2.  When focussing on LSD Stokes $V$ profiles only, we assume a Doppler width \vD\ of the local profile equal to $\vD=5.3$~\kms\ 
\citep[as in][]{Donati23} so that synthetic Stokes $I$ profiles are as wide as the observed ones given the assumed \vsini\ (see Table.~\ref{tab:par}) and minimal Zeeman 
broadening.  When simultaneously fitting LSD Stokes $IV$ or Stokes $IVQU$ profiles, we assume a more realistic Doppler width of $\vD=3.5$~\kms, yielding a profile width matching 
the asymptotic width of LSD profiles from atomic lines of decreasing Land\'e factors \citep[see][]{Donati23}.  

We also introduce the polarization filling factor $f_P$ (assumed 
constant over the whole star) describing the fraction of each grid cell that contributes to the large-scale field and to Stokes $V$, $Q$ and $U$ profiles, implying a magnetic field 
of $B_P/f_P$ within the magnetic portion of a cell and a magnetic flux of $B_P$ over the whole cell.  Similarly, we assume that a fraction $f_I$ of each grid cell (called 
filling factor of the small-scale field, again equal for all cells) hosts small-scale fields of strength $B_P/f_P$, implying a small-scale magnetic flux over the whole cell equal 
to $B_I = B_P f_I/f_P$.  This simple parametric approach within ZDI is used to empirically reproduce the co-existence of small-scale and large-scale fields at the surface of M 
dwarfs, as inferred from both observations and simulations \citep[e.g.,][]{Kochukhov21,Yadav15}.  To compute the synthetic light curve, we sum the photometric contributions 
estimated from the local brightness and limb angle of all visible cells at each observed rotation phase.  
Finally, we assume that the projection of the stellar rotation axis on the plane of the sky makes an angle $\psi$ with respect to the reference direction 
of the polarimeter (i.e., east-west), counting anti-clockwise.  The values of $i$ and \vsini\ for AU~Mic and EV~Lac are adopted from previous studies (see Table~\ref{tab:par}).  
{\emr None of the parameters describing the local line profile, and in particular \vD, $f_P$ and $f_I$, are assumed to depend on the magnetic field. } 

We stress that one challenge of the third approach is to realistically model the observed LSD Stokes $IQUV$ profiles, which are essentially weighted averages of about 1500 real spectral 
lines with different strengths, Zeeman sensitivities and patterns, with a single virtual line described as a simple Zeeman triplet.  Whereas this approach is rather straightforward 
in the case of stars with weak magnetic fields, its implementation is more tricky for strongly magnetic stars like AU~Mic and EV~Lac, requiring a careful adjustment of line 
parameters and in particular of \vD, $f_P$ and $f_I$ for the model to match observations.  
It was already shown to be successful at simultaneously reproducing sets of observed LSD Stokes $IV$ profiles in the case of AU~Mic \citep{Donati23}, and suggested that 
the inferred large-scale field is significantly underestimated in strongly magnetic stars when focusing on Stokes $V$ profiles alone \citep[as in, e.g.,][]{Klein21,Bellotti24}.  
Generalizing this approach to the simultaneous modelling of LSD Stokes $IVQU$ profiles makes the tuning of parameters even more essential.  We nonetheless succeeded in achieving 
a reasonably good fit to full Stokes spectropolarimetric observations of both AU~Mic and EV~Lac, as described in the following paragraphs.  
The alternative approach described in 
\citet{Rosen15}, based on constructing tables of synthetic LSD profiles for a wide variety of field strengths and orientations from synthetic polarized spectra, is hardly applicable 
at nIR wavelengths where synthetic spectra significantly disagree between different sets of atmospheric models \citep[e.g.,][]{Cristofari22a} and do not provide a reliable 
description of the spectral content of high-resolution observations of M dwarfs \citep{Artigau18,Artigau24} except in narrow regions where line parameters are manually adjusted 
\citep[e.g.,][]{Cristofari23}.  

When fitting LSD Stokes $IVQU$ profiles, we only used the Stokes $I$ profiles associated with the Stokes $V$ ones and not the (mostly redundant) ones associated with Stokes $QU$ profiles 
to avoid putting excessive weight on Stokes $I$ vs Stokes $VQU$ data.  

\begin{figure*}[ht!]
\center{\includegraphics[scale=0.31,angle=-90]{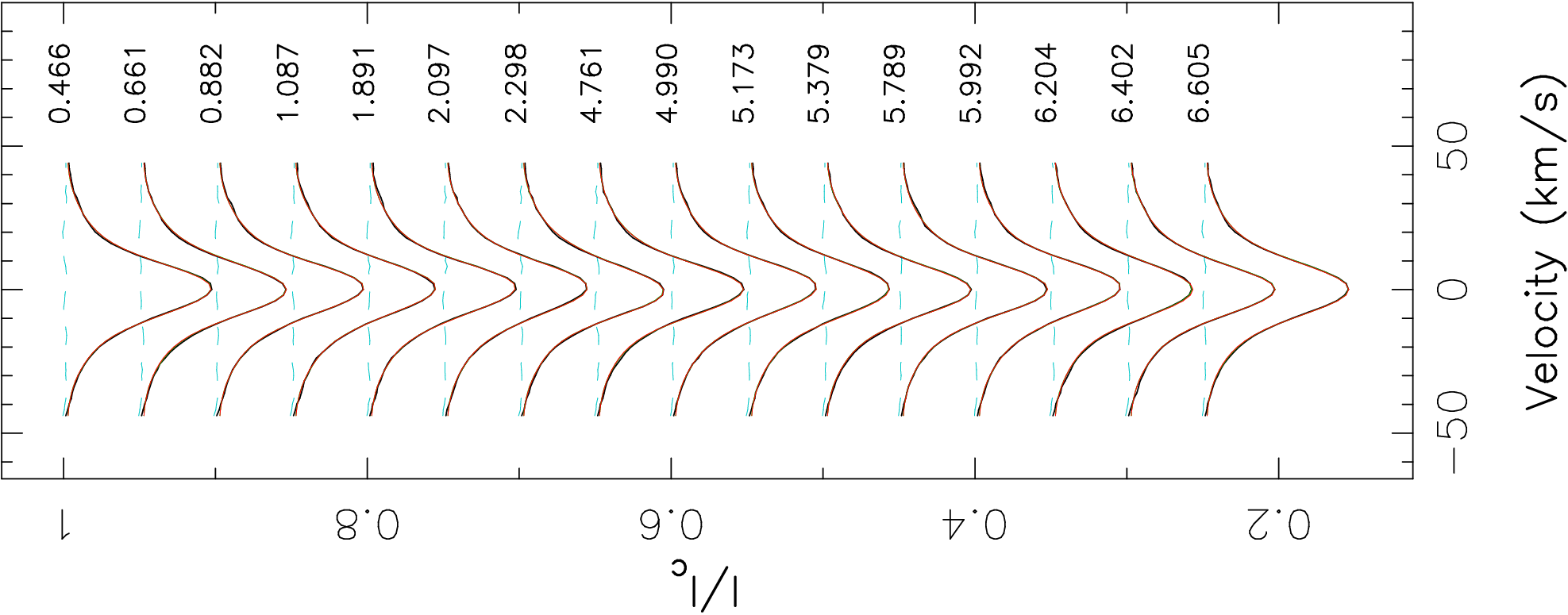}\hspace{2mm}\includegraphics[scale=0.31,angle=-90]{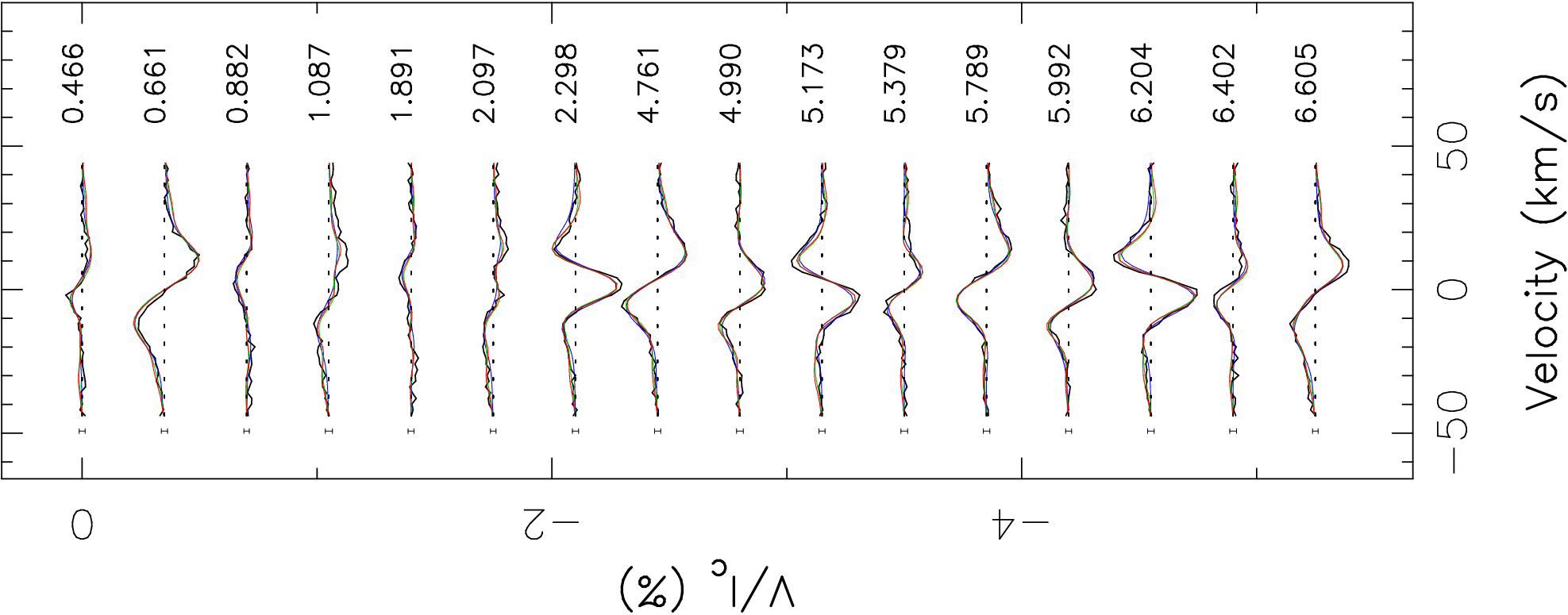}\hspace{2mm}\includegraphics[scale=0.31,angle=-90]{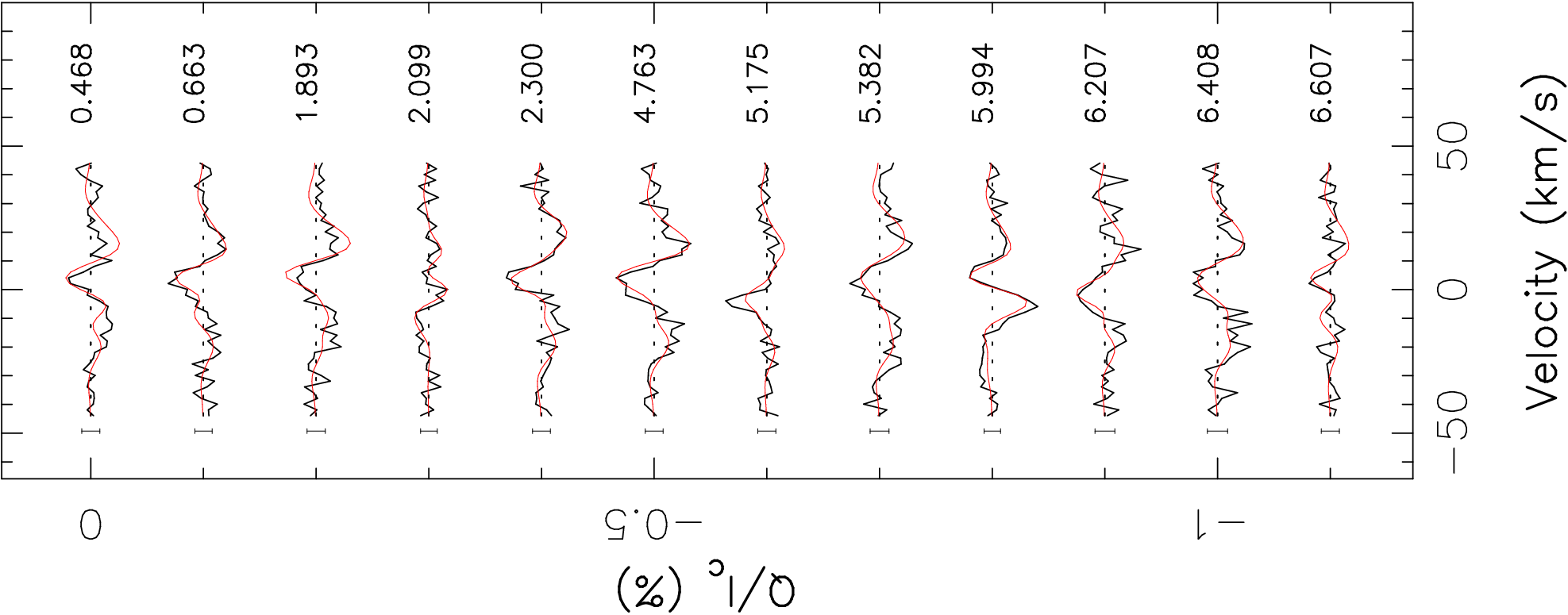}\hspace{2mm}\includegraphics[scale=0.31,angle=-90]{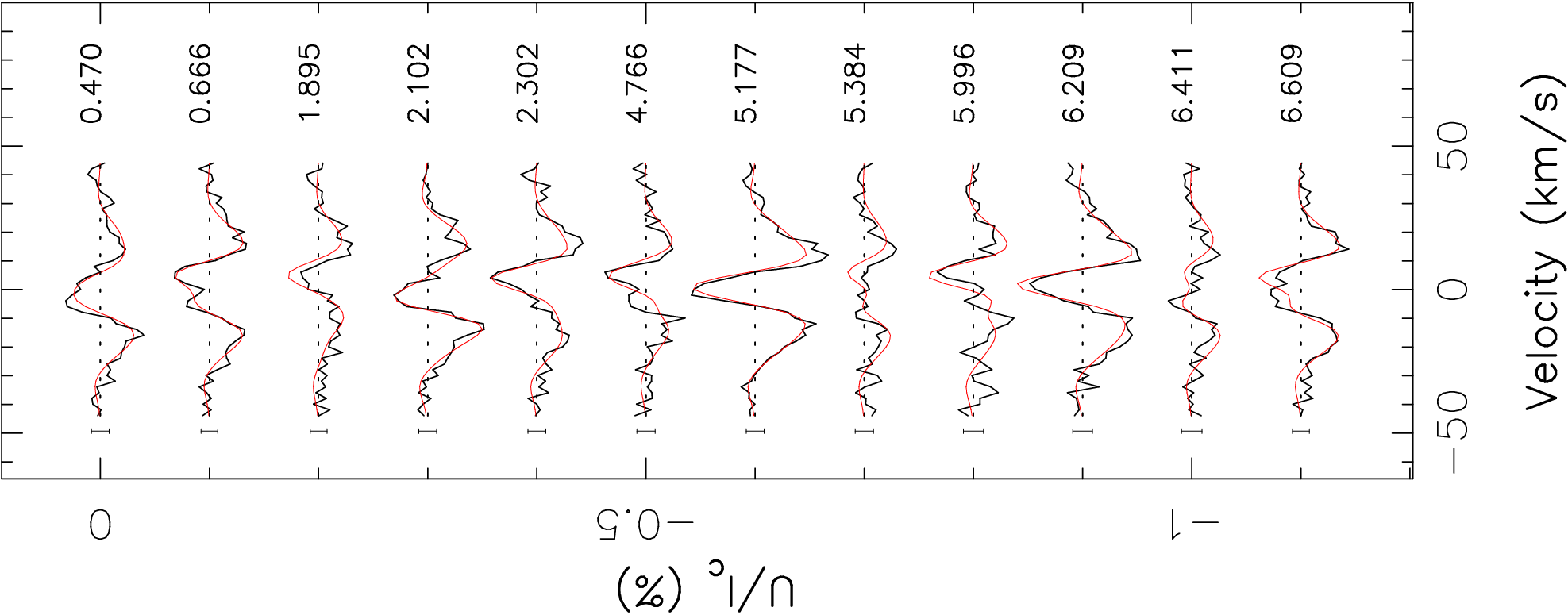}\vspace{1mm}}
\center{\includegraphics[scale=0.31,angle=-90]{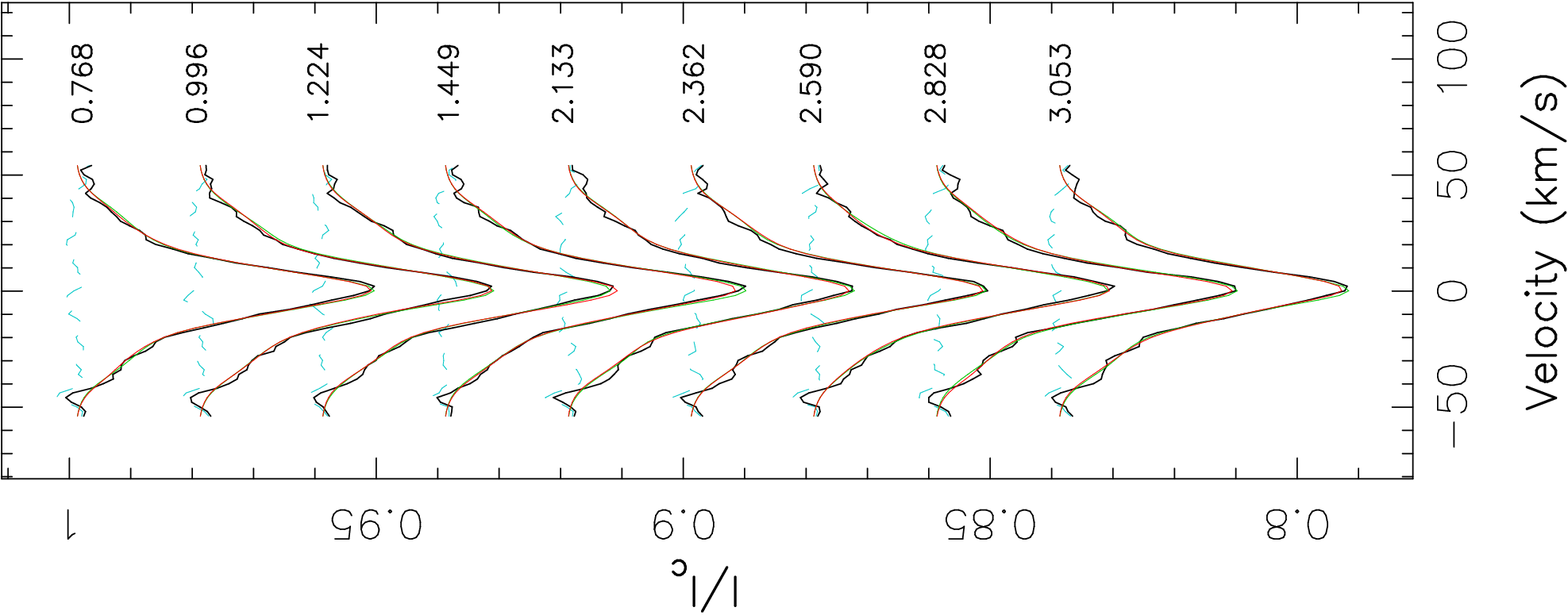}\hspace{2mm}\includegraphics[scale=0.31,angle=-90]{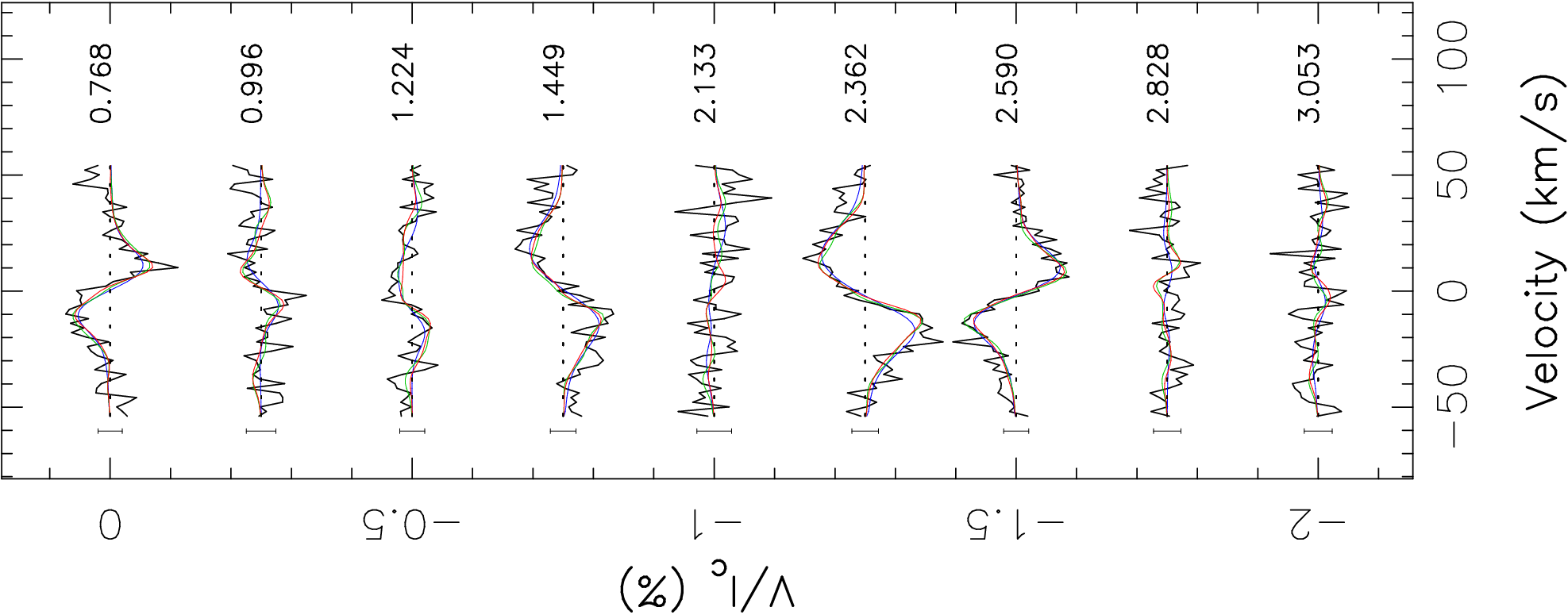}\hspace{2mm}\includegraphics[scale=0.31,angle=-90]{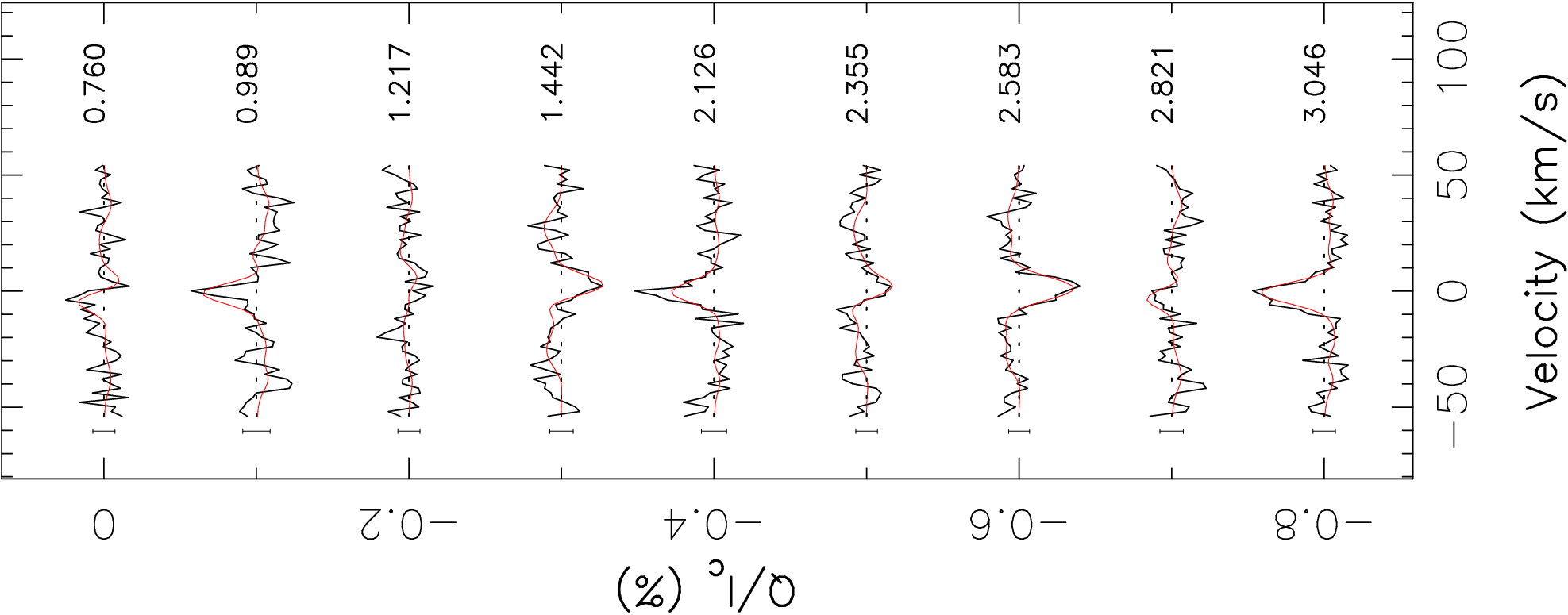}\hspace{2mm}\includegraphics[scale=0.31,angle=-90]{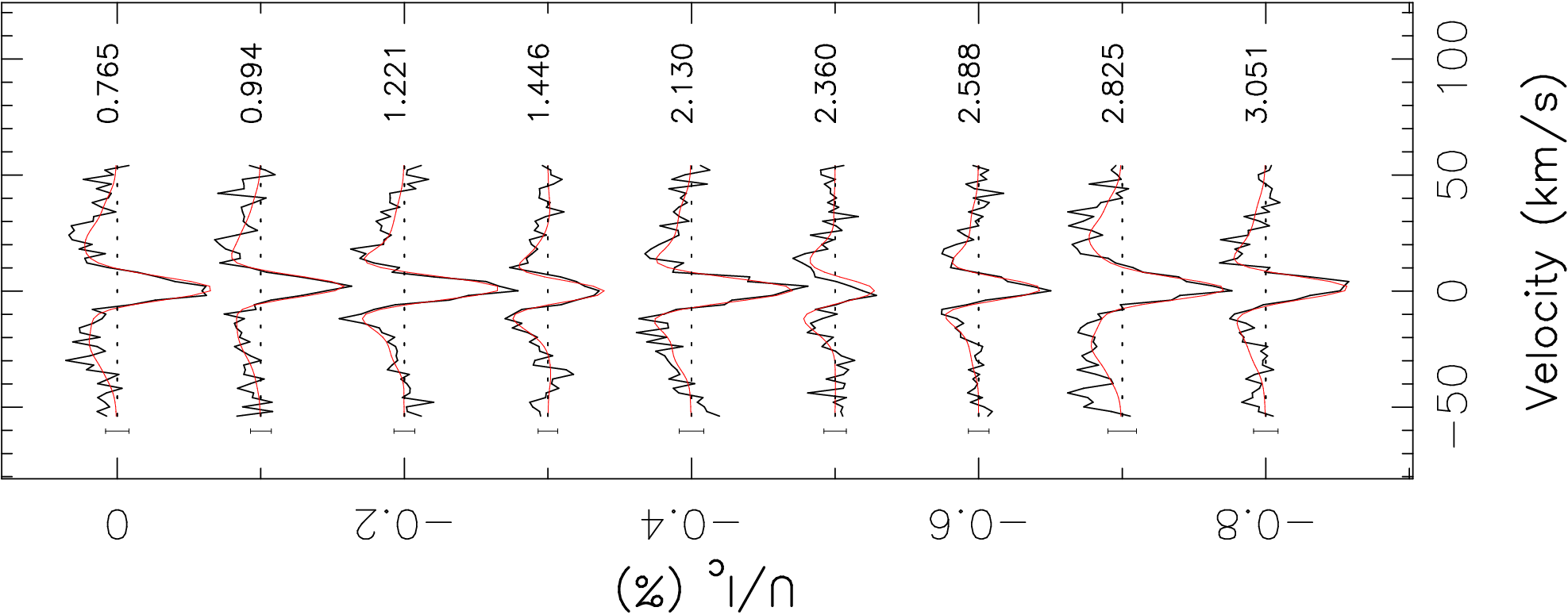}} 
\caption[]{Observed (thick black line) and modelled (thin color line) LSD Stokes $I$, $V$, $Q$ and $U$ profiles (from left to right) of AU~Mic (top row) and EV~Lac (bottom row).  
The thin blue, green and red lines respectively correspond to the Stokes $V$, Stokes $IV$ and Stokes $IVQU$ ZDI fits to the LSD profiles, whereas the cyan dashed lines in the left 
panels illustrate the difference between the observations and the Stokes $IVQU$ ZDI fit.  
Rotation cycles are indicated to the right of LSD profiles, and $\pm$1$\sigma$ error bars are added to the left of Stokes $VQU$ signatures. }
\label{fig:fit}
\end{figure*}

\begin{figure*}[ht!]
\flushright{\large\textbf{Stokes \textit{V}}   \raisebox{0.3\totalheight}{\includegraphics[scale=0.44,angle=-90]{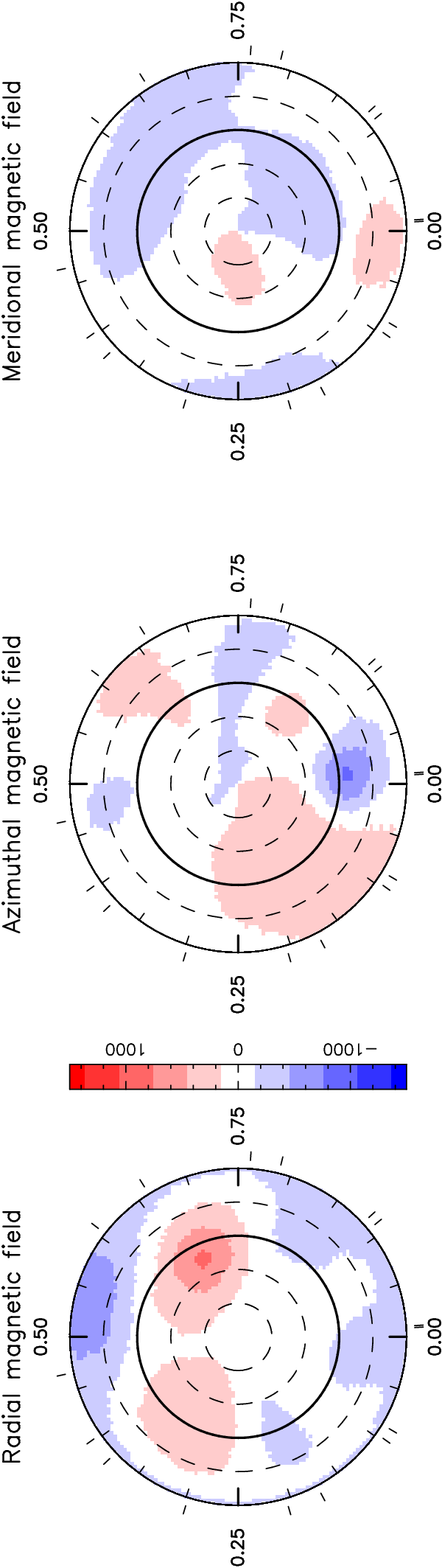}}\vspace{1mm}}
\flushright{\large\textbf{Stokes \textit{IV}}  \raisebox{0.3\totalheight}{\includegraphics[scale=0.44,angle=-90]{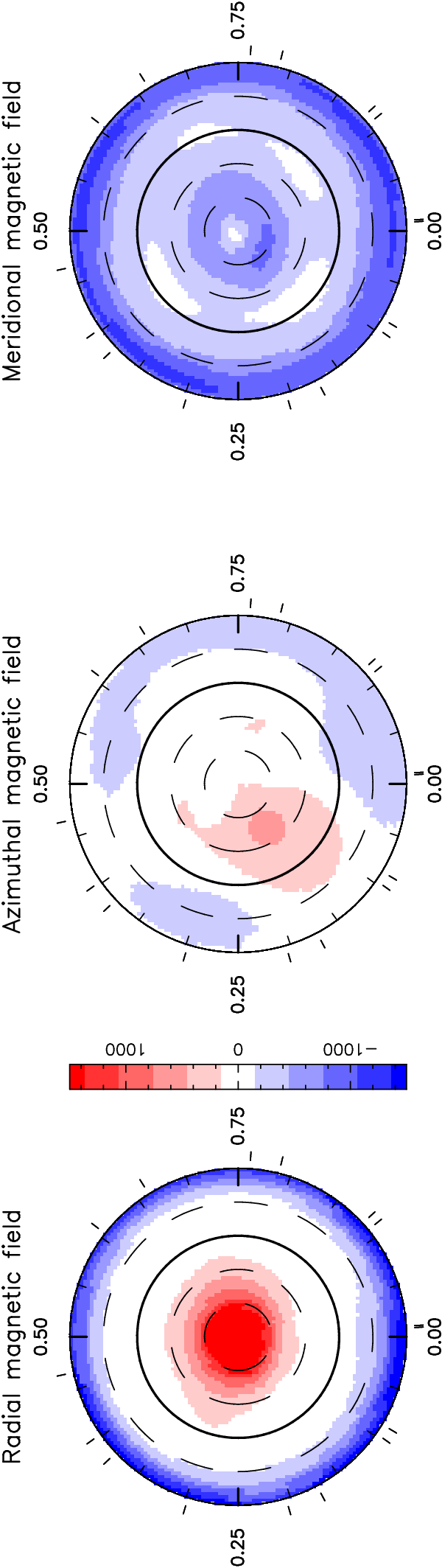}}\vspace{1mm}}
\flushright{\large\textbf{Stokes \textit{IVQU}}\raisebox{0.3\totalheight}{\includegraphics[scale=0.44,angle=-90]{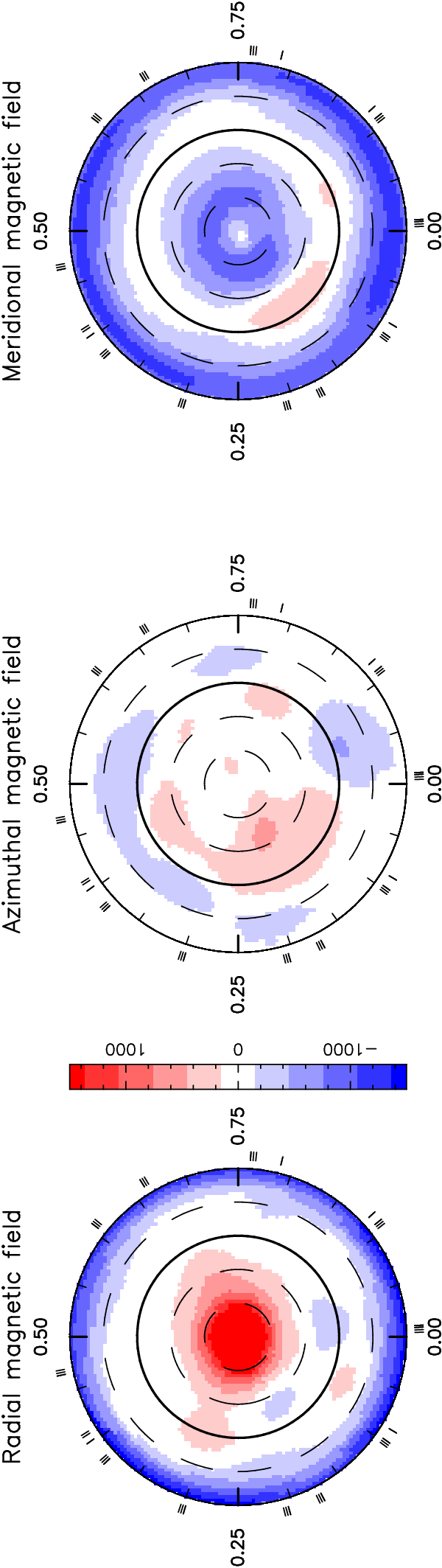}}}
\caption[]{Reconstructed maps of the large-scale magnetic field of AU~Mic using ZDI from Stokes $V$ (top row), Stokes $IV$ (middle row) and Stokes $IVQU$ (bottom row) 
SPIRou data.  In each row and from left to right, we show the radial, azimuthal and meridional components of the magnetic field in spherical coordinates (in G).
All maps are shown in a flattened polar projection down to latitude $-60$\degr, with the north pole at the center and the equator depicted as a bold line.
Outer ticks indicate phases of observations.  Positive radial, azimuthal and meridional fields respectively point outwards, counterclockwise and polewards.  }
\label{fig:map-au}
\end{figure*}

\begin{figure*}[ht!]
\flushright{\large\textbf{Stokes \textit{V}}   \raisebox{0.3\totalheight}{\includegraphics[scale=0.44,angle=-90]{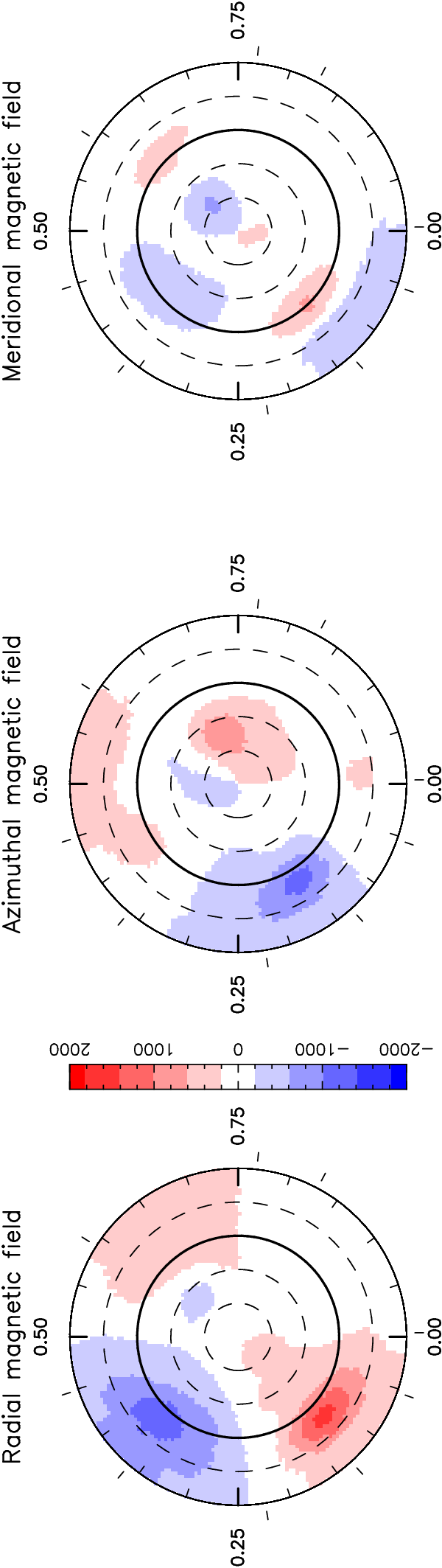}}\vspace{1mm}}
\flushright{\large\textbf{Stokes \textit{IV}}  \raisebox{0.3\totalheight}{\includegraphics[scale=0.44,angle=-90]{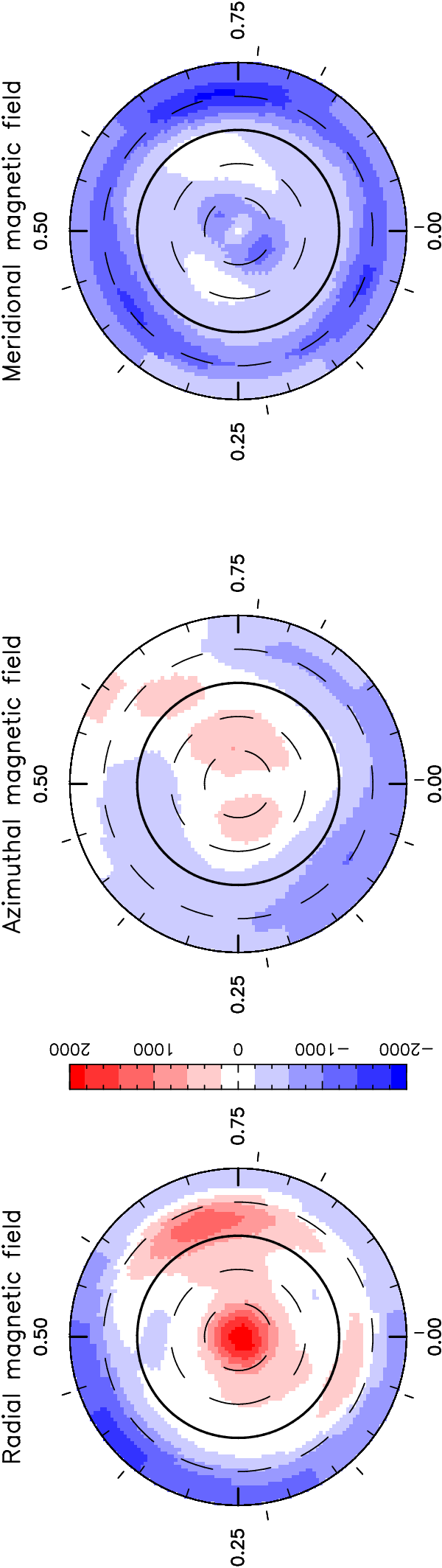}}\vspace{1mm}}
\flushright{\large\textbf{Stokes \textit{IVQU}}\raisebox{0.3\totalheight}{\includegraphics[scale=0.44,angle=-90]{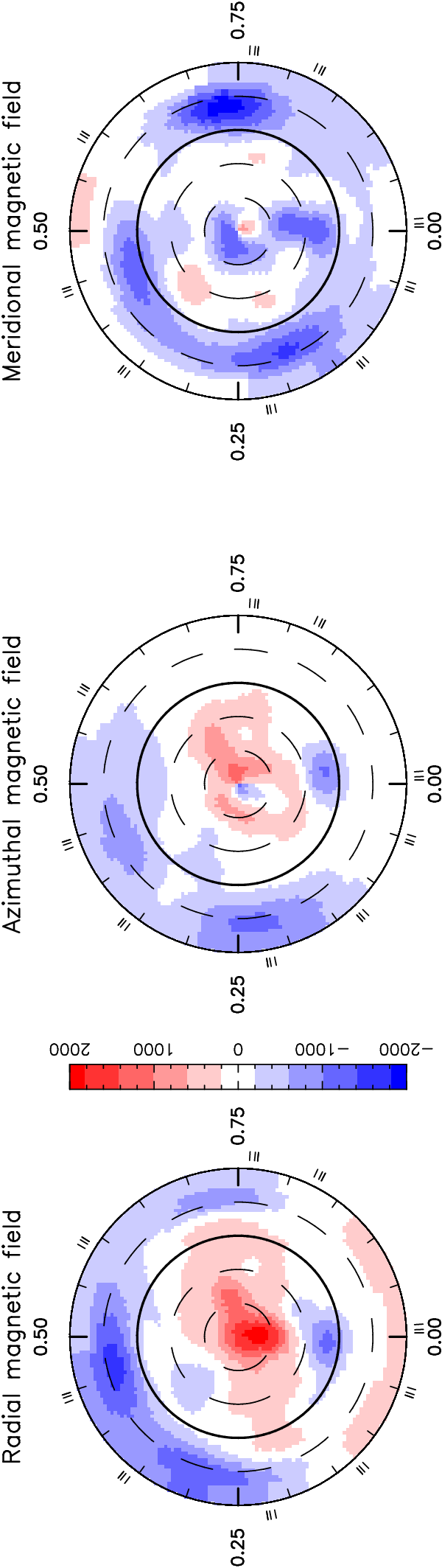}}}
\caption[]{Same as Fig.~\ref{fig:map-au} for EV~Lac. } 
\label{fig:map-ev}
\end{figure*}

\subsection{Results for AU~Mic and EV~Lac}

\begin{figure*}[ht!]
\center{\includegraphics[scale=0.32,angle=-90]{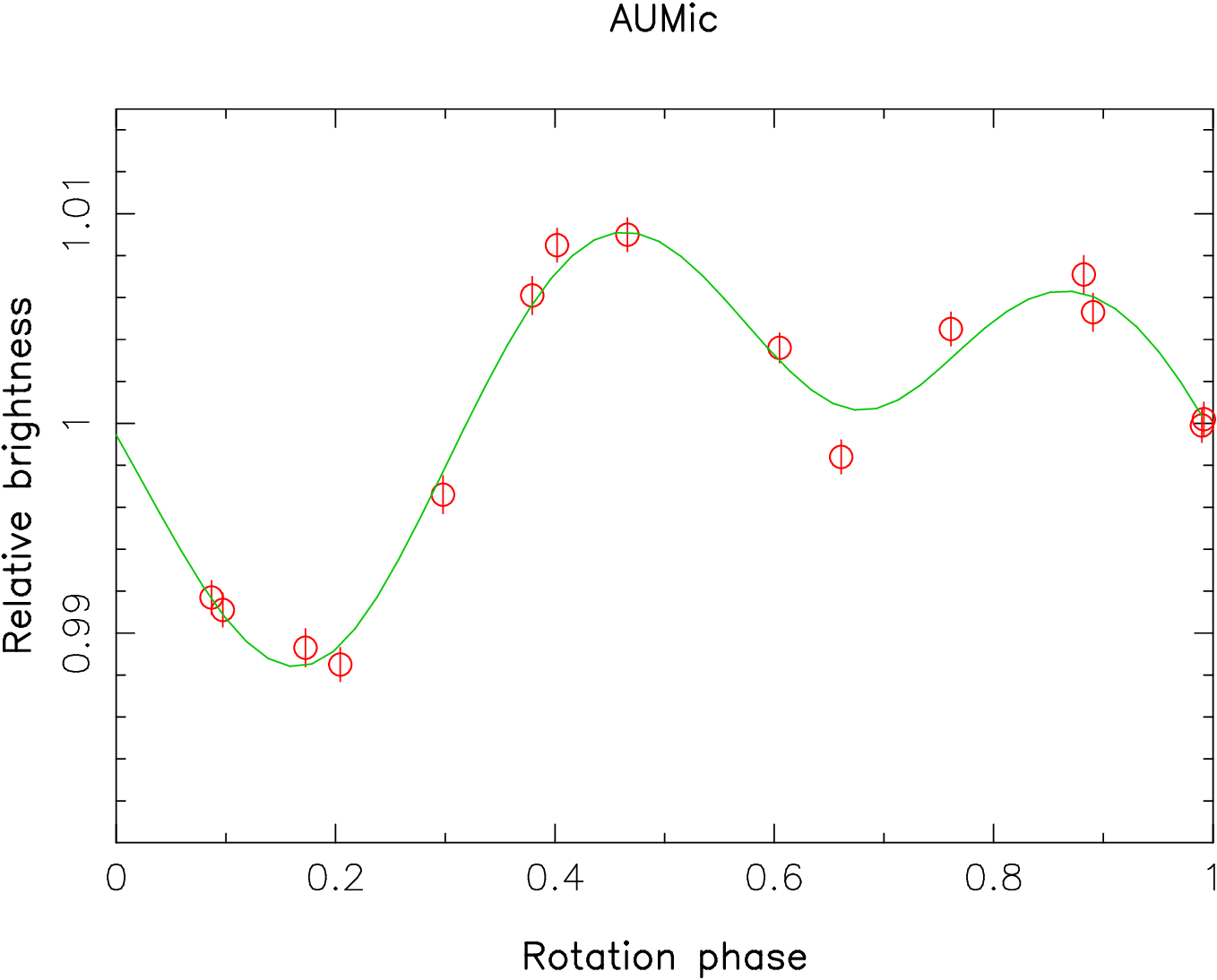}\hspace{5mm}\includegraphics[scale=0.45,angle=-90]{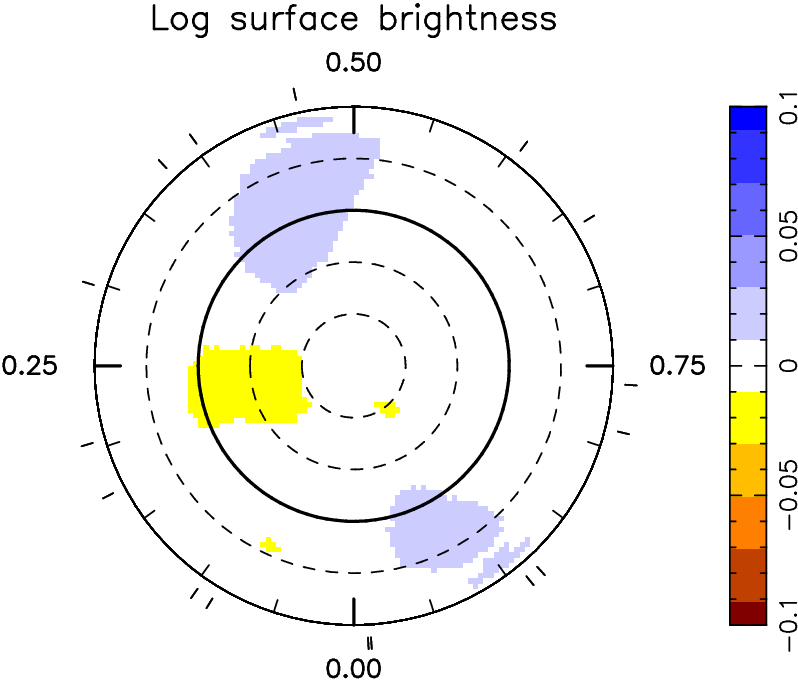}} 
\center{\includegraphics[scale=0.32,angle=-90]{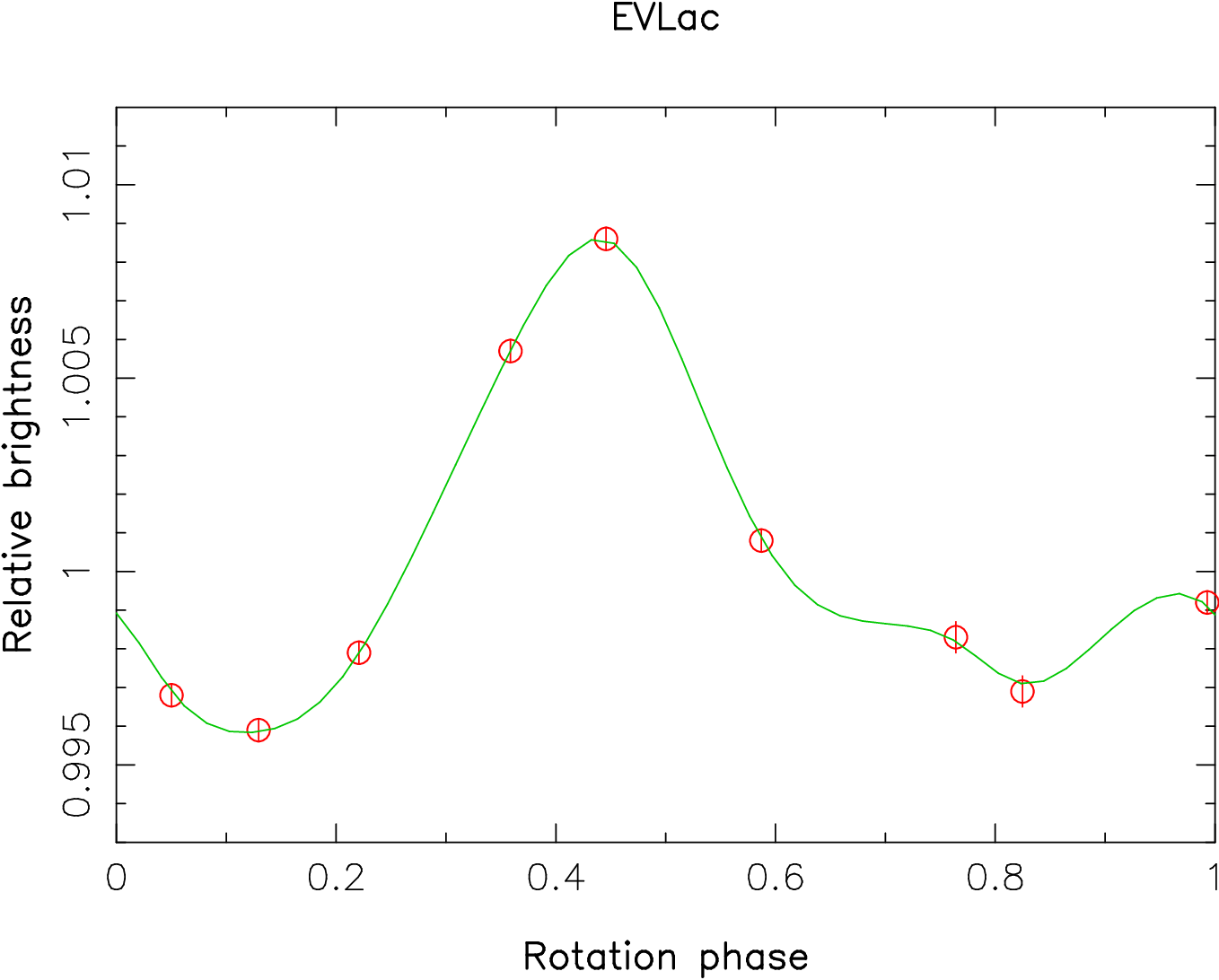}\hspace{5mm}\includegraphics[scale=0.45,angle=-90]{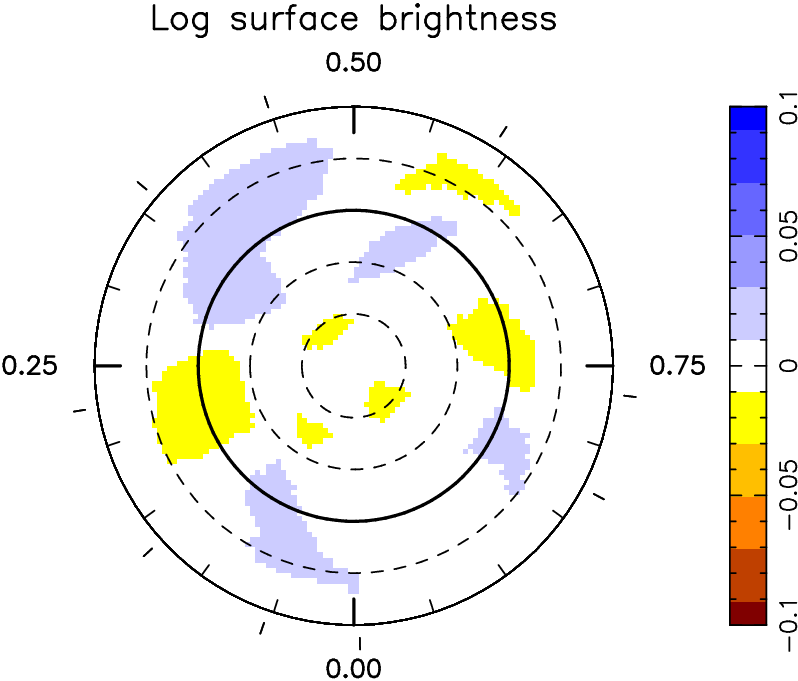}} 
\caption[]{Photometric light curves inferred from $dT$ estimates (red circles, left panels) with ZDI fits (green curves), and brightness maps (right panels) reconstructed simultaneously 
with the magnetic maps of Figs.~\ref{fig:map-au} and \ref{fig:map-ev}, for both AU~Mic (top row) and EV~Lac (bottom row), in the case of Stokes $IVQU$ reconstructions.  
(Very similar results are obtained in Stokes $IV$ reconstructions).  
The brightness maps are shown as in Figs.~\ref{fig:map-au} and \ref{fig:map-ev}, with yellow and blue depicting regions cooler and warmer than the quiet photosphere, respectively.  } 
\label{fig:pho}
\end{figure*}

We show in Fig.~\ref{fig:fit} the observed LSD Stokes $IVQU$ profiles of both stars along with the corresponding ZDI fits to the data down to photon noise level for each of the three 
imaging cases mentioned above.  The best results were obtained for $f_P=0.2$ and $f_I=0.9$ for AU~Mic, in agreement with \citet{Cristofari23} and \citet{Donati23}, and for $f_P=0.15$ and 
$f_I=1.0$ for EV~Lac \citep[with $f_I$ values matching][]{Cristofari23,Cristofari25}.  The three ZDI fits to the LSD Stokes $V$ profiles and the two ZDI fits to the LSD Stokes $I$ profiles (depicted 
with different colors in the first- and second-column plots of Fig.~\ref{fig:fit}) are of very similar quality.  The derived ZDI images are shown in Fig.~\ref{fig:map-au} for AU~Mic and 
Fig.~\ref{fig:map-ev} for EV~Lac.  Reconstructed simultaneously with ZDI along with magnetic maps, brightness images are found to only show low contrast features as expected from the small 
amplitude of photometric curves (see Fig.~\ref{fig:pho}).  

\begin{figure*}[ht!]
\center{\includegraphics[scale=0.32,angle=-90]{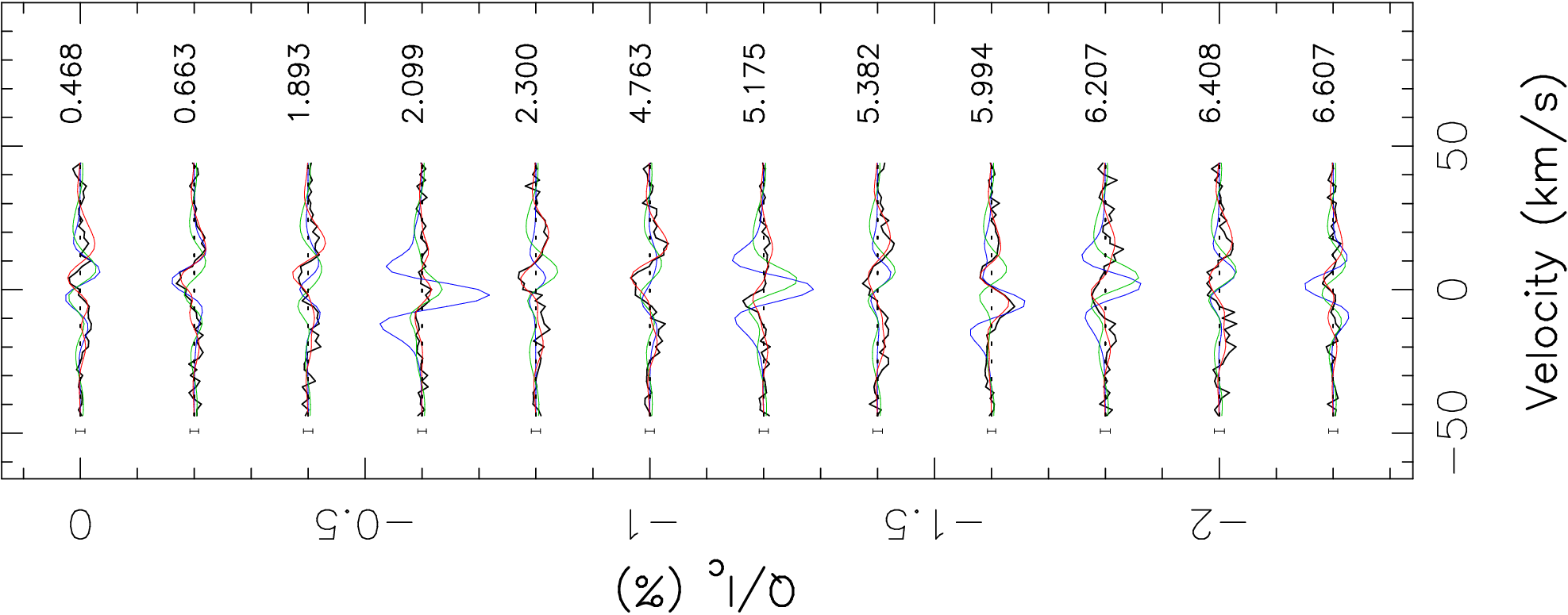}\hspace{2mm}\includegraphics[scale=0.32,angle=-90]{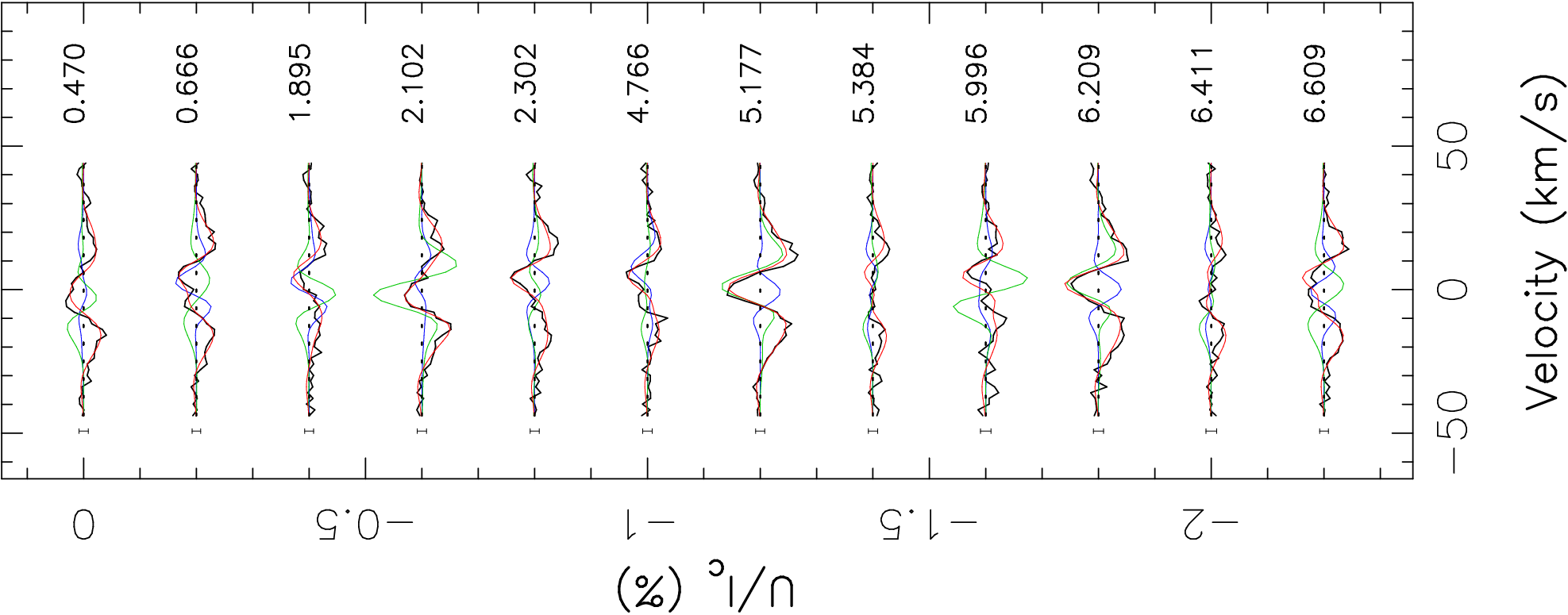}\hspace{5mm}\includegraphics[scale=0.32,angle=-90]{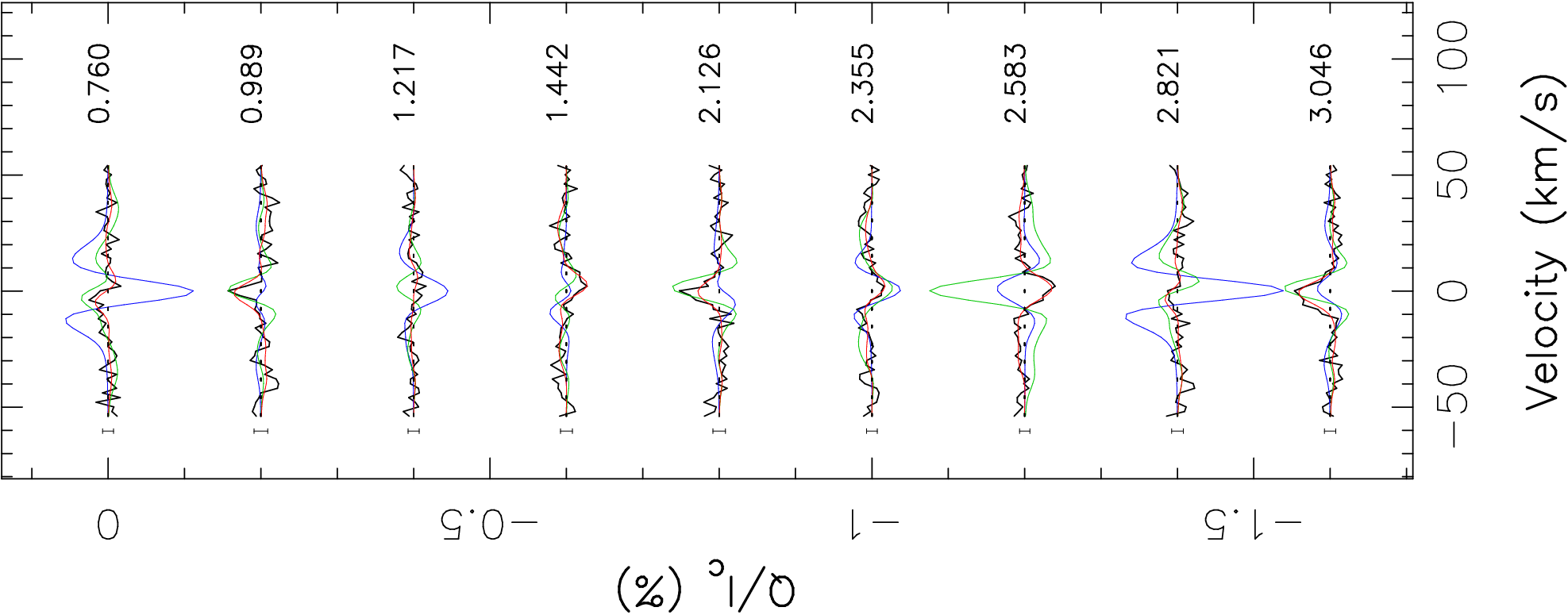}\hspace{2mm}\includegraphics[scale=0.32,angle=-90]{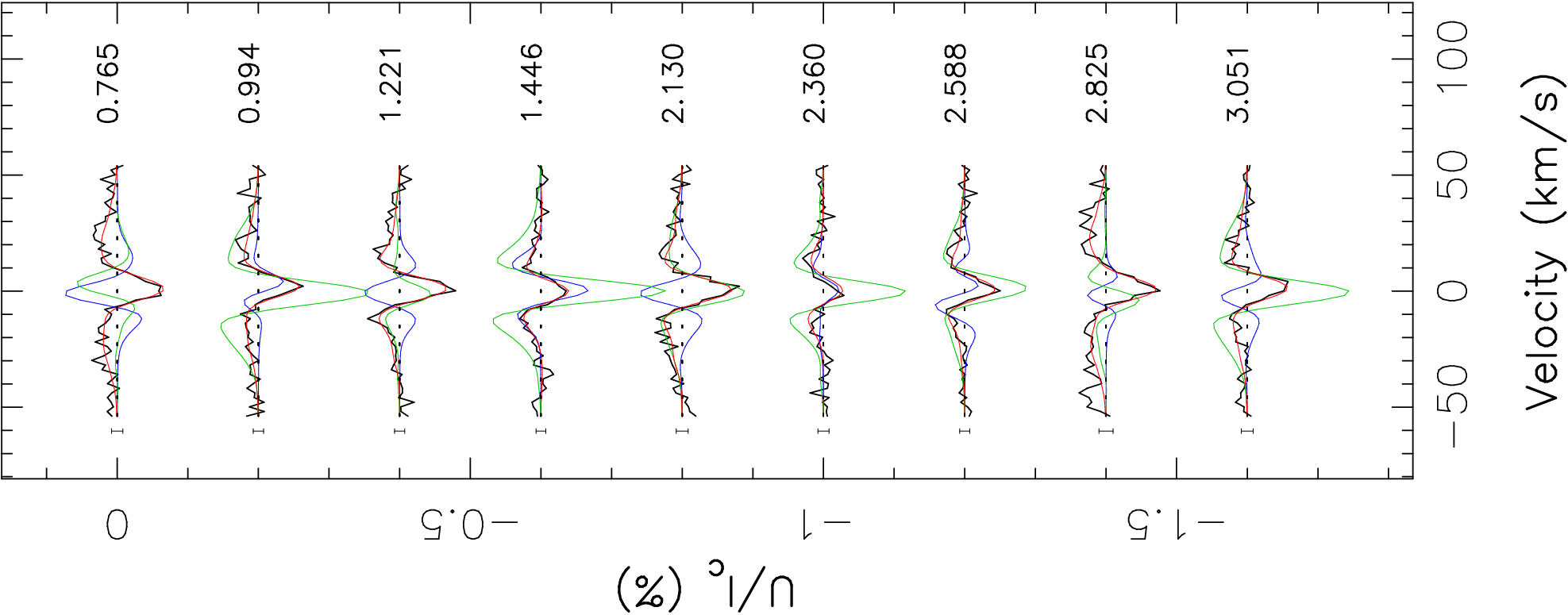}}
\caption[]{Same as the two rightmost panels of Fig.~\ref{fig:fit}, showing now also the synthetic Stokes $QU$ profiles associated with the Stokes $V$ (blue line) and $IV$ (green line) 
reconstructions as a comparison, for both AU~Mic (left panels) and EV~Lac (right panels). } 
\label{fig:fit2}
\end{figure*}

Regarding AU~Mic, we find that the reconstructed large-scale field is dominantly poloidal in all three reconstructions (see Table~\ref{tab:mag} where the main magnetic properties of the 
derived fields are summarized).  When considering the LSD Stokes $V$ profiles alone (and setting $\vD=5.3$~\kms, see Sec.~\ref{sec:met}), the inferred field is mostly non-axisymmetric, 
with a global dipole field of only 250~G (see Table~\ref{tab:mag}) and weaker quadrupole and octupole components.  The small-scale field derived with ZDI in this initial case (1.6~kG) is significantly 
weaker than that obtained in the two other imaging cases (4~kG), reflecting that the Zeeman broadening of the LSD Stokes $I$ profiles requires stronger fields when setting $\vD=3.5$~\kms.  In our 
modeling context where we assume that the small-scale field grows with the large-scale field (with a scaling factor $f_I/f_P$), it implies that we also miss a significant fraction of the large-scale field when 
fitting the LSD Stokes $V$ profiles only, as already concluded in \citet{Donati23}.  This can indeed easily happen as a result of equatorial symmetry in cases like AU~Mic where the rotation 
axis of the star is nearly perpendicular to the line of sight, causing the Stokes $V$ signatures of odd axisymmetric terms in the SH expansion of the magnetic field to mostly cancel out.  The 
large-scale field that we derive when adjusting both LSD Stokes $I$ and $V$ profiles is almost 3$\times$ stronger, exceeding 0.9~kG, with a global dipole field of about 1.1~kG and a (mostly 
aligned) octupole field of about 0.8~kG, each storing 70 and 26~percent of the overall magnetic energy reconstructed in the (dominant) poloidal field component.  These components 
are stronger in our new reconstruction than in those of our previous paper \citep{Donati23}, as a likely result of both the intrinsic temporal variability and the smearing that ZDI induces 
by handling Stokes $IV$ data collected over a time span of over six months.  The two low-latitude 
positive radial field features reconstructed in the first imaging case at phases 0.35 and 0.65 now only show up as weak appendages of the main positive radial field feature straddling the 
pole in the derived Stokes $IV$ image.  Adding in the LSD Stokes $QU$ profiles in the ZDI fit enhances these appendages, but otherwise generates only moderate changes to the reconstructed 
magnetic topology and its main characteristics (see Table~\ref{tab:mag}).  We stress that the synthetic $QU$ profiles associated with the Stokes $V$ and Stokes $IV$ reconstructions are 
often discrepant in amplitude and shape with those of our full Stokes modelling (see Fig.~\ref{fig:fit2}, left panels) even though the Stokes $IV$ and Stokes $IVQU$ magnetic maps are not that different, 
emphasizing the importance of collecting Stokes $QU$ data whenever possible.  Besides, it demonstrates that our implementation of ZDI from all LSD Stokes parameters behaves satisfactorily, 
reliably reproducing LSD Stokes $QU$ profiles with only small changes from magnetic topologies derived from LSD Stokes $IV$ profiles, for stars like AU~Mic whose large-scale field is not 
too complex and mostly axisymmetric.  

\begin{table}[ht!]
\caption[]{Properties of the large-scale magnetic field of AU~Mic and EV~Lac, for the Stokes $V$, Stokes $IV$ and Stokes $IVQU$ ZDI reconstruction schemes} 
\centering 
\resizebox{\linewidth}{!}{  
\begin{tabular}{ccccccccc}
\hline
Quantity             && \multicolumn{3}{c}{AU~Mic} && \multicolumn{3}{c}{EV~Lac} \\
                     && $V$ & $IV$ & $IVQU$        && $V$ & $IV$ & $IVQU$        \\ 
\hline
\vD\ (\kms)           && 5.3 & 3.5  & 3.5           && 5.3  & 3.5  & 3.5   \\ 
$f_P$                && 0.2 & 0.2  & 0.2           && 0.15 & 0.15 & 0.15  \\ 
$f_I$                &&     & 0.9  & 0.9           &&      & 1.0  & 1.0  \\ 
$\psi$ (\degr)       &&     &      & $26\pm5$      &&      &      & $86\pm5$ \\ 
<$B_P$> (G)          && 355 & 935  & 920           && 520 & 995 & 845 \\ 
<$B_I$> (kG)         &&  1.6 &  4.2 & 4.1          &&  3.5 &  6.6 &  5.6 \\ 
<$B_{I,\rm max}$> (kG) &&  3.5 & 9.5 & 10.1         && 10.3 & 13.2 & 12.5 \\ 
<$B_s$> (kG)         &&  1.6 &  2.6 & 2.6          &&  2.5 &  4.3 &  4.2 \\ 
\Bd\ (kG)            && 0.25 &  1.15 &  1.06         && 0.34 & 0.80 & 0.62 \\  
tilt (\degr)         && 34   & 11   & 9           && 90   & 28   & 35   \\ 
phase                && 0.54 & 0.54 & 0.57         && 0.92 & 0.82 & 0.90 \\ 
poloidal (percent)   && 84   & 98   & 97           && 95   & 89   & 82   \\ 
axisymetry (percent) && 29   & 97   & 96           && 2    & 82   & 54   \\ 
pol dipole (percent) && 37 & 70 & 60          && 32   & 63   & 45   \\ 
\hline 
\end{tabular}}
\tablefoot{\emr In each case we list the assumed Doppler width \vD\ and polarization $f_P$ filling factor used in the ZDI modeling, the inferred values of the angle $\psi$ of the projected stellar rotation axis
on the plane of the sky, the reconstructed large-scale field strength <$B_P$> and corresponding average small-scale field <$B_I$> quadratically averaged over the whole star,
the maximum small-scale field <$B_{\rm max}$> at the surface of the star, the phase-averaged small-scale field <$B_s$> computed over the visible (limb darkened) stellar hemisphere
(directly comparable to the small-scale field <$B$> measured with ZeeTurbo), the polar strength \Bd\ of the large-scale field
dipole component, the tilt of the dipole field to the rotation axis and the phase towards which it is tilted, the amount of magnetic energy reconstructed in the poloidal
component of the large-scale field, and in the axisymmetric dipole mode ($m=0$, $\ell=1$) of this component.  Error bars are typically equal to 10~percent for field
strengths and percentages and 5\degr\ for field inclinations. }
\label{tab:mag}
\end{table}

\begin{figure*}[ht!]
\center{\includegraphics[scale=0.37,angle=-90]{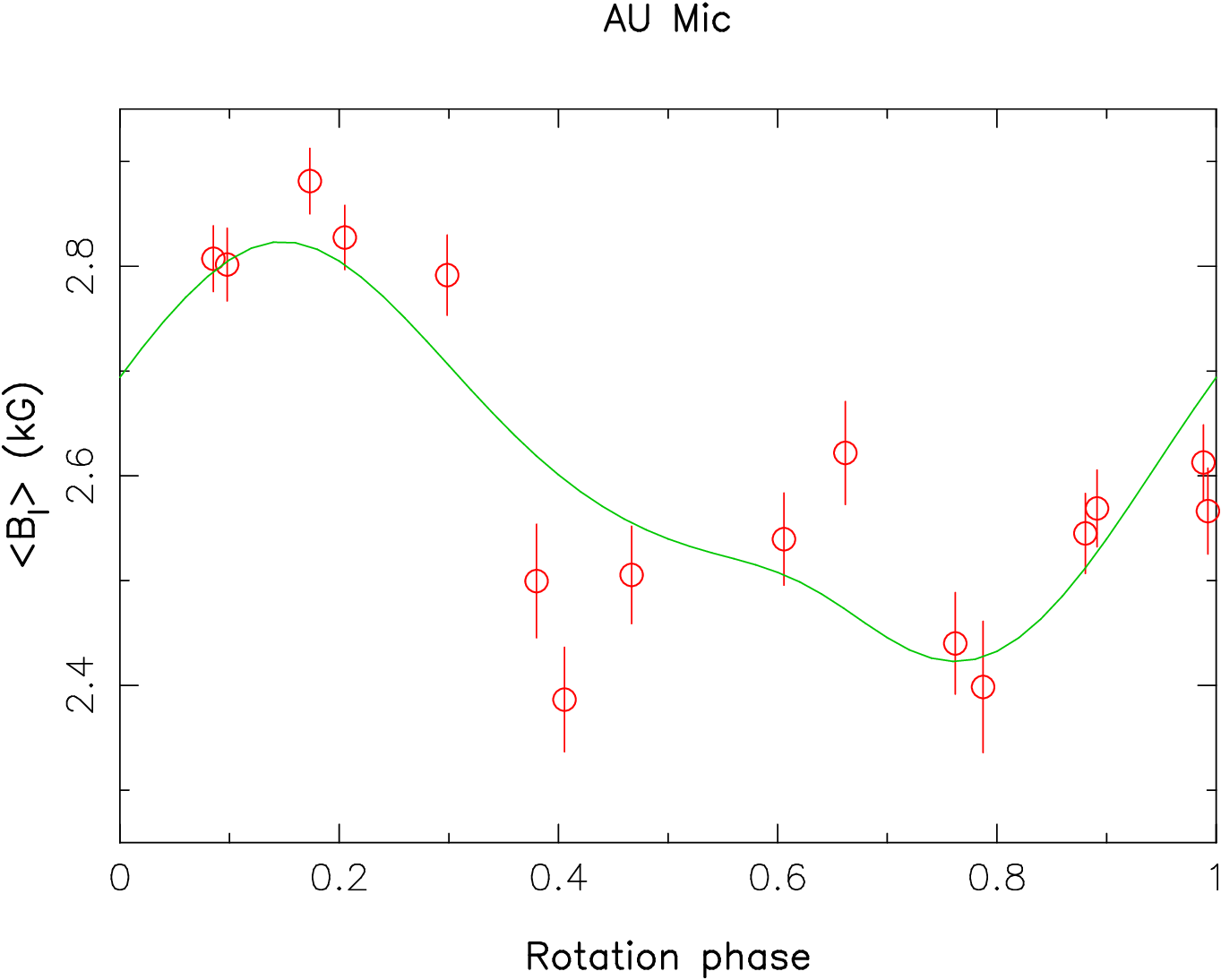}\hspace{2mm}\includegraphics[scale=0.37,angle=-90]{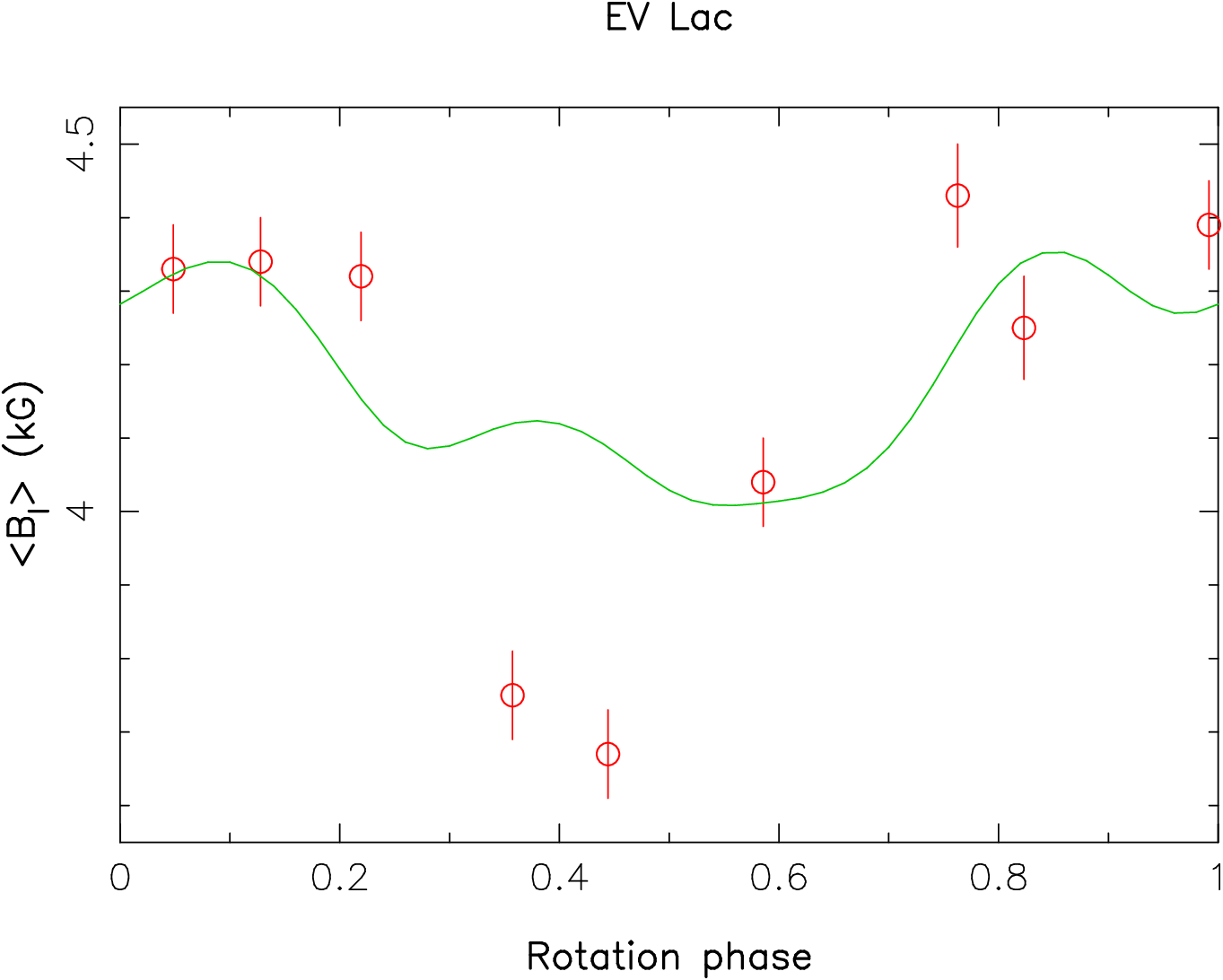}} 
\caption[]{Rotational modulation of the small-scale field <$B_I$> (green curve), computed over the visible (limb darkened) stellar hemisphere and derived from our ZDI reconstructions of AU~Mic (left panel) 
and EV~Lac (right panel), versus that of the actual average small-scale field <$B_s$> directly measured from individual Stokes $I$ spectra with ZeeTurbo \citep[red circles,][]{Cristofari23,Cristofari25}.} 
\label{fig:bmod}
\end{figure*}

Regarding EV~Lac, we also obtain that the reconstructed large-scale field is largely poloidal in all three cases.  The mostly non-axisymmetric large-scale field reconstructed from the LSD 
Stokes $V$ profiles alone does not properly account for the large Zeeman broadening of the Stokes $I$ profiles, in particular in the far wings, even after increasing the scaling factor $f_I/f_P$ 
between small-scale and large-scale fields by 50 percent from the AU~Mic case (to better match the shape of the strongest observed LSD Stokes $V$ profiles, and in particular their gradients 
close to the line center).  The reconstructed large-scale field includes a dominant quadrupole term storing 40 percent of the poloidal field energy and generating 0.6~kG equatorial regions of 
alternating polarities, a weaker (0.34~kG) highly tilted dipole component, and a weak and complex octupole field.  The inferred phase-averaged small-scale field <$B_s$> computed over the (limb-darkened) 
visible stellar hemisphere (2.5~kG) is significantly weaker than the one measured 
with ZeeTurbo.  Incorporating the LSD Stokes $I$ profiles in the ZDI fit (while reducing \vD\ from 5.3 to 3.5~\kms) yields a fairly different field topology, with the 
radial field component now featuring a strong positive field region near the pole accompanied by lower latitude appendages, but still exhibiting a similar polarity pattern with phase  
(positive radial field at phases 0.1 and 0.6 and negative field at phase 0.4).  In this process, both the average large- and small-scale magnetic strengths grow by about 50~percent, respectively 
reaching about 1.0 and 6.6~kG.  The global dipole component increases as well to 0.80~kG and is now dominant, being no longer orthogonal to the rotation axis but still significantly tilted (by about 30\degr).  
Finally, adding in LSD Stokes $QU$ profiles to the ZDI fit further complexifies the field, with the main field characteristics showing in particular that the dipole field decreases (to 0.62~kG) but 
remains dominant (see Table~\ref{tab:mag}).  Some power is now present in the higher SH modes, e.g., with a negative radial field region showing up at low latitudes near phase 0.0.  
As for AU~Mic and even more so, the synthetic $QU$ profiles associated with the Stokes $V$ and Stokes $IV$ reconstructions are often grossly off in both amplitude and shape with those of our full Stokes 
modelling (see Fig.~\ref{fig:fit2}, right panels), emphasising again that LSD Stokes $QU$ profiles can provide key information about the large-scale and small-scale magnetic topologies, especially for 
low-\vsini\ stars like EV~Lac whose narrow spectral lines contain limited spatial information about stellar surface features.  

{\emr We caution that, although our ZDI reconstruction of EV~Lac is made unique by construction (i.e., by choosing the simplest solution consistent with the data for the selected parameters, see Sec.~\ref{sec:met}), 
the inversion problem is still degenerate to some level in cases where the rotational broadening, the spectral resolution, the number of observed profiles and the SNR are limited, even when all Stokes profiles 
are included in the process.  We suspect that different, even more complex magnetic topologies, may also exist for EV~Lac, matching equally well the LSD Stokes $IVQU$ data.  The situation is better for AU~Mic 
where the larger \vsini\ stabilizes the inversion and yields a more constrained solution.} 

To further assess the consistency of our ZDI reconstructions from all Stokes profiles, we computed for both stars the small-scale fields that we expect from the derived magnetic topologies, 
as if measured from the spectra over the visible stellar hemispheres, as well as their rotational modulation (see Fig.~\ref{fig:bmod}).  In both cases, we find small-scale field values 
and modulation that are reasonably consistent with ZeeTurbo measurements, in the range 2.4-2.9~kG for AU~Mic and 3.7-4.4~kG for EV~Lac \citep[see Tables~\ref{tab:log-au} 
and \ref{tab:log-ev} and][]{Cristofari23,Donati23,Cristofari25}, even though the modulation amplitude we recover with ZDI for EV~Lac is smaller than what ZeeTurbo indicates 
(missing the dip at phase 0.4).  It confirms in particular that reconstructions from LSD Stokes $IV$ and $IVQU$ profiles yield a far better match to 
observations than those from LSD Stokes $V$ profiles alone.  

\begin{figure*}[ht!]
\center{\includegraphics[scale=0.24,angle=-90]{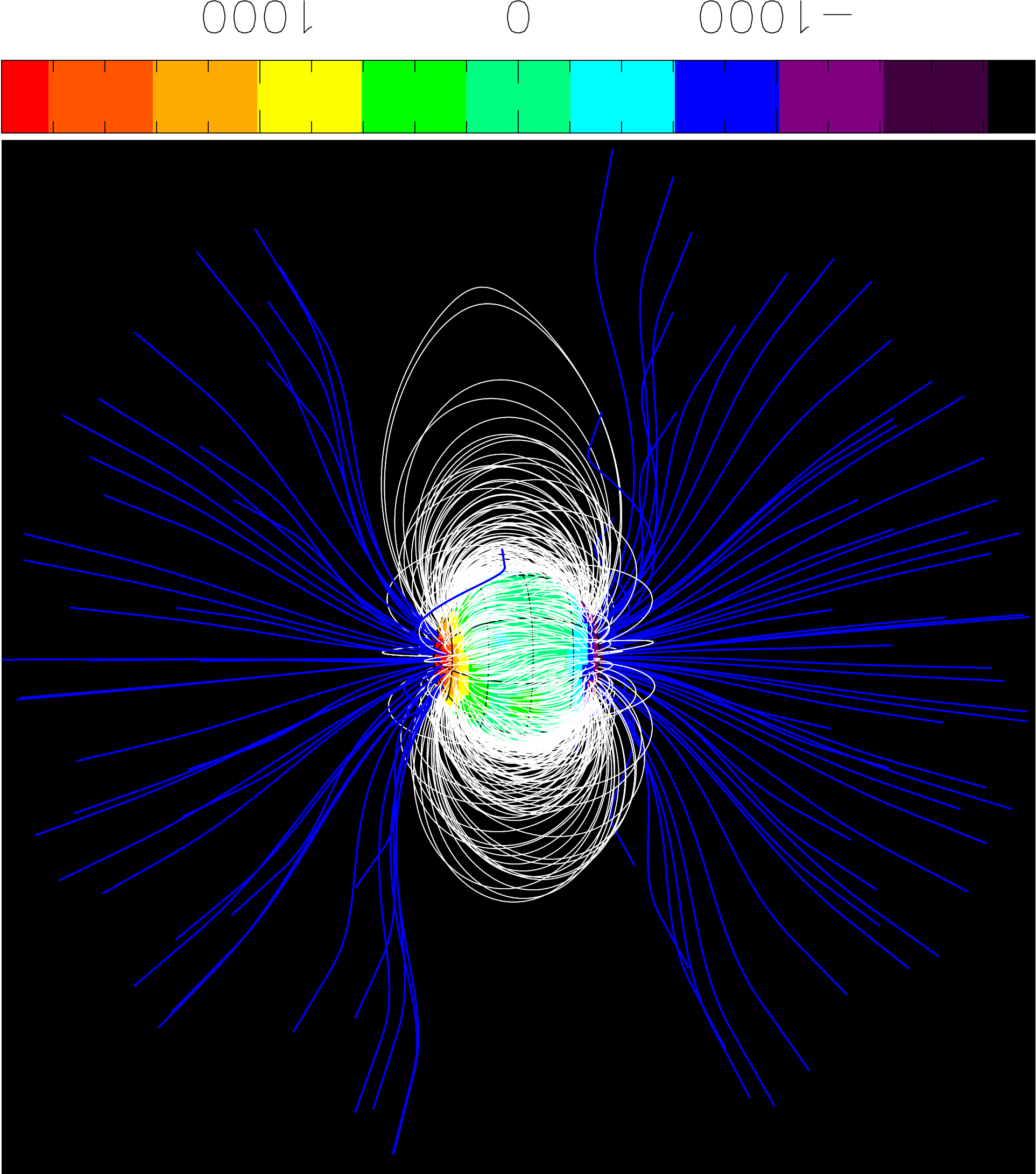}\hspace{2mm}\includegraphics[scale=0.24,angle=-90]{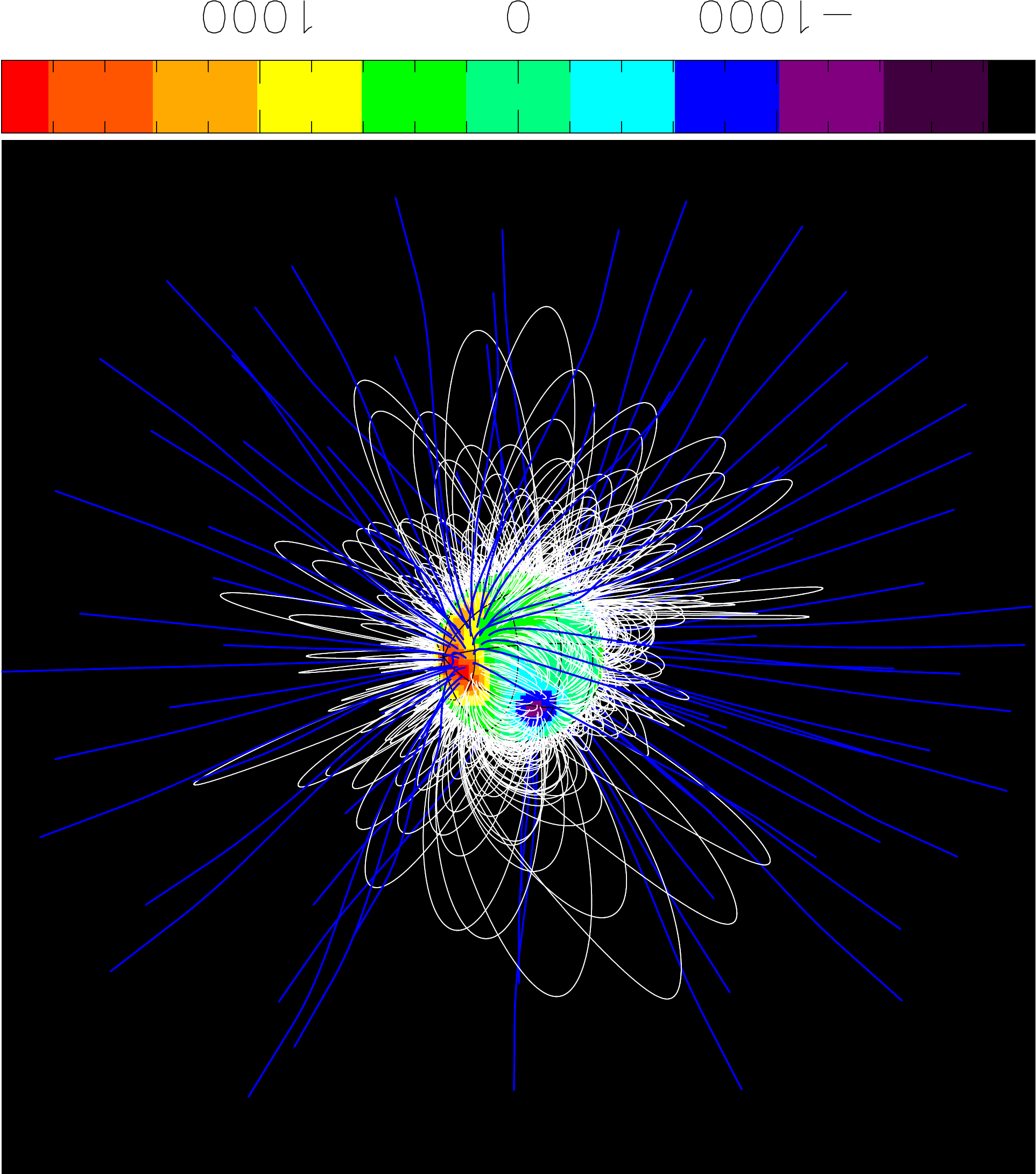}} 
\caption[]{Potential field extrapolation of the large-scale field topologies derived with ZDI from our SPIRou Stokes $IVQU$ observations of AU~Mic (left panel) and EV~Lac 
(right panel), with the open field lines shown in blue and the closed field ones in white.  The color scale depicts the radial field value at the surface of the star (in G), 
as derived by ZDI.  The stars are shown as viewed from the Earth at rotational phases 0.20 and 0.90 respectively, with the source surface arbitrarily set to 5~\rstar.  The 
color scale depicts the radial field value at the surface of the star (in G), as derived by ZDI.  Animated movies can be found at 
\url{https://spirou-legacy.irap.omp.eu/lib/exe/fetch.php?media=jfd:aumic-23.gif} and \url{https://spirou-legacy.irap.omp.eu/lib/exe/fetch.php?media=jfd:evlac-23.gif}. }  
\label{fig:extr}
\end{figure*}

We also derived potential field extrapolations of the large-scale field of both stars as seen from an Earth-based observer, using the radial field maps derived from all LSD Stokes profiles 
(see Fig.~\ref{fig:extr}) and assuming a source surface located at 5~\rstar.  

\section{Summary, discussion and conclusion}
\label{sec:dis}

In this paper we presented a spectropolarimetric analysis of the well-known active M dwarfs AU~Mic and EV~Lac, from a set of 67 circularly and linearly polarized spectra collected on both 
stars with SPIRou at CFHT from August to October 2023 and covering a few rotational cycles.  We first used LSD to derive the Zeeman signatures associated with these spectra from about 1500 atomic 
lines, and detected the weak LSD Stokes $QU$ signals of both stars at most observed phases in addition to the much stronger Stokes $IV$ signals usually recorded on M dwarfs with SPIRou.  
We then employed ZDI to reconstruct the magnetic topologies at the surface of both stars, using either LSD Stokes $V$ data alone, or LSD Stokes $IV$ profiles, or finally all LSD Stokes $IVQU$ 
observations, to investigate in detail how the inferred topologies differ from one another and what additional information each type of data brings to the inversion problem and helps reducing 
potential degeneracies.  In the second and third cases, we further constrained ZDI by simultaneously reconstructing surface brightness distributions, fitting photometric light curves derived 
from $dT$ measurements at the same time as LSD Stokes profiles.  

The magnetic imaging process and ZDI routines used in this study are the same as in \citet{Donati23}, which we now also applied on data sets including LSD Stokes $QU$ profiles.  
When running ZDI to LSD Stokes $V$ profiles only \citep[and setting $\vD=5.3$~\kms\ to match the average width of LSD Stokes $I$ profiles, as in][]{Donati23}, the inferred large-scale magnetic 
topology of both stars is mostly non-axisymmetric, and the quadratically averaged large-scale and small-scale fields over the stellar surfaces only reach 355~G and 1.6~kG for AU~Mic, 
and 520~G and 3.5~kG for EV~Lac.  When adjusting LSD Stokes $I$ and $V$ profiles simultaneously (setting now $\vD=3.5$~\kms), the large-scale and small-scale fields both increase to about 935~G 
and 4.2~kG for AU~Mic, and 1.0~kG and 6.6~kG for EV~Lac, with the large-scale field being now dominantly axisymmetric and incorporating a stronger dipole component, of about 1.1~kG and 0.8~kG 
in the case of AU~Mic and EV~Lac.  When fitting all LSD Stokes profiles at once, the large-scale and small-scale fields evolve again, though less so for AU~Mic than for EV~Lac.  
Our results also show that our simple parametric approach at describing both the large-scale and small-scale fields at the surface of the star within ZDI is successful at 
reproducing observations in a consistent way.  {\emr We however caution that, given the low \vsini\ and the limited SNR and spatial information enclosed in spectral lines, the inversion problem is 
still ill-posed, in the case of EV~Lac in particular, even for reconstructions from Stokes $IVQU$ data, and that more complex solutions may still exist, matching equally all LSD profiles.}   

The magnetic topology we recover for AU~Mic from all Stokes profiles is fairly simple, with almost all of the recovered magnetic energy stored in the dipole and octupole field components, both mostly 
aligned with the rotation axis and thus perpendicular to the line of sight.  {\emr We do not expect the nearly equator-on orientation of AU~Mic to be the main cause for the reconstructed largely 
axisymmetric magnetic topology, especially when all Stokes profiles are fitted, the larger \vsini\ allowing non-axisymmetric structures to be more reliably recovered. } 
The large-scale magnetic topology of AU~Mic is conveniently oriented for Earth-based observers to detect and monitor 
radio emission, in particular the reported highly circularly polarized and rotationally modulated bursts of coherent beamed emission generated by the electron cyclotron maser instability likely 
taking place in the auroral rings of the magnetic poles \citep{Bloot24}.  Given the stronger large-scale dipole field derived in our new study, one can expect a wider Alfven volume than that initially 
estimated \citep{Kavanagh21,Alvarado22}, implying that the two transiting planets (b and c) are likely located within this volume and thus able to magnetically interact with their host star.  However, the observed 
radio bursts from AU~Mic have not shown so far periodicity at the orbital or the synodic periods of the planets, over the radio monitoring campaign of \citet{Bloot24}, suggesting that the transiting 
planets do not induce beamed radio emission as a result of star-planet interactions, or that the corresponding radio flux is weaker than expected.  It nonetheless sets AU~Mic as an ideal laboratory for 
studying radio emission from the electron cyclotron maser instability, either that caused by activity or induced by close-in planets, for objects other than the Sun and Jupiter.  

The magnetic topology we infer for EV~Lac from all Stokes profiles is stronger, more complex and more tilted with respect to the rotation axis, than that of AU~Mic, with a weaker (though still 
dominant) dipole component and significant power in higher terms of the SH expansion, inducing smaller magnetic features at the stellar surface.  Our result shows in particular that the large-scale 
field of EV~Lac and its dipole component are not highly tilted to the rotation axis, as initially suggested from magnetic maps reconstructed from Stokes $V$ data only \citep{Morin08b,Bellotti24}.  
{\emr Despite their residual non axisymmetry, our new magnetic maps of EV~Lac are much more consistent with the magnetic topologies of other M dwarfs with similar internal structure (like AU~Mic or AD~Leo)}.  
The more complex magnetic geometry of EV~Lac versus that of AU~Mic may be a hint at why EV~Lac exhibits one of the highest fractions of flaring time among a sample of active M dwarfs \citep[including AU~Mic,][]{Paudel24}:  
the more complex the magnetic topology, the smaller the size of closed field loops at the surface of the star, the more frequent flaring is likely to happen as a result of reconnection triggered by surface 
convection and differential rotation.  
More such reconstructions of magnetic topologies of M dwarfs, derived from rotationally modulated sets of LSD Stokes $IVQU$ (or at least Stokes $IV$) profiles, are needed to further 
document how magnetic complexity depends on stellar parameters, whereas further dynamo simulations are required at the same time to explore these aspects in more depth.  

Our results suggest that magnetometric studies of active stars with kG fields, including T~Tauri stars, benefit from spectropolarimetric observations with SPIRou carried out in 
all Stokes parameters, provided the required SNR can be reached for Stokes $QU$ Zeeman signatures to be detected, so that their magnetic topologies can be reconstructed more reliably, 
especially for stars with medium \vsini's for which the Zeeman distortions of spectral lines are not drowned into rotational broadening.  Further improvement should also come from gaining 
simultaneous spectropolarimetric access to both optical and nIR domains so that spectral line distortions from magnetic fields can be disentangled from those induced by brightness 
features, and both be consistently reconstructed with ZDI from sets of rotationally modulated line profiles \citep[as in][]{Donati24}.  This should become possible at CFHT once VISION, the 
forthcoming interface allowing both ESPaDOnS and SPIRou to operate at the same time with no loss of polarimetric efficiency, is implemented at the telescope within a couple of semesters from now.

\section*{Data availability}  SPIRou data used in this study will be publicly available from 01 August 2025 at the Canadian Astronomy Data Center (\url{https://www.cadc-ccda.hia-iha.nrc-cnrc.gc.ca}).       

\begin{acknowledgements}
This work benefited from the SIMBAD CDS database at URL {\tt http://simbad.u-strasbg.fr/simbad} and the ADS system at URL {\tt https://ui.adsabs.harvard.edu}.  
Our study is based on data obtained at the CFHT, operated by the CNRC (Canada), INSU/CNRS (France) and the University of Hawaii.
The authors wish to recognise and acknowledge the very significant cultural role and reverence that the summit of Maunakea has always had
within the indigenous Hawaiian community.  
\end{acknowledgements}

\bibliographystyle{aa}
\bibliography{aa55428-25}
\clearpage

\begin{appendix}

\section{Observation log}
\label{sec:appA}

Tables~\ref{tab:log-au} and \ref{tab:log-ev} gives the full log of Stokes $VQU$ spectropolarimetric SPIRou observations of AU~Mic and EV~Lac from August to October 2023.   

\begin{table*}[ht!]
% \small
\caption[]{Observing log of our Stokes $VQU$ spectropolarimetric SPIRou observations of AU~Mic in August 2023} 
\centering 
\resizebox{\linewidth}{!}{  
\begin{tabular}{ccccccccccc}
\hline
BJD        & UT date & Star & c / $\phi$ & t$_{\rm exp}$ & SNR & $\sigma_P$            & Stokes & \Bl\  &  <$B_s$> &  $dT$ \\
(2459000+) &         &      &            &   (s)        & ($H$) & ($10^{-4} I_c$)      &        &  (G)  &  (kG)    &  (K)  \\
\hline
1158.9450165 & 02 Aug 2023 & AU~Mic & 238 / 0.466 & 802.4 & 834 & 0.81 & $V$ & 5$\pm$4 & 2.51$\pm$0.05 & 19.19$\pm$1.18 \\
1158.9557904 & 02 Aug 2023 & AU~Mic & 238 / 0.468 & 802.4 & 845 & 0.63 & $Q$ & & & \\
1158.9664817 & 02 Aug 2023 & AU~Mic & 238 / 0.470 & 802.4 & 845 & 0.64 & $U$ & & & \\
1159.8927438 & 03 Aug 2023 & AU~Mic & 238 / 0.661 & 802.4 & 889 & 0.80 & $V$ & 93$\pm$4 & 2.62$\pm$0.05 & 3.70$\pm$1.14 \\
1159.9035273 & 03 Aug 2023 & AU~Mic & 238 / 0.663 & 802.4 & 876 & 0.61 & $Q$ & & & \\
1159.9143589 & 03 Aug 2023 & AU~Mic & 238 / 0.666 & 802.4 & 895 & 0.61 & $U$ & & & \\
1160.9676210 & 04 Aug 2023 & AU~Mic & 238 / 0.882 & 802.4 & 920 & 0.74 & $V$ & -13$\pm$4 & 2.54$\pm$0.04 & 16.45$\pm$1.31 \\
1161.9622558 & 05 Aug 2023 & AU~Mic & 239 / 0.087 & 802.4 & 893 & 0.95 & $V$ & 76$\pm$5 & 2.81$\pm$0.03 & -6.09$\pm$1.21 \\
1165.8684823 & 09 Aug 2023 & AU~Mic & 239 / 0.891 & 802.4 & 908 & 0.80 & $V$ & -12$\pm$4 & 2.57$\pm$0.04 & 13.82$\pm$1.26 \\
1165.8795421 & 09 Aug 2023 & AU~Mic & 239 / 0.893 & 802.4 & 925 & 0.64 & $Q$ & & & \\
1165.8903909 & 09 Aug 2023 & AU~Mic & 239 / 0.895 & 802.4 & 923 & 0.61 & $U$ & & & \\
1166.8725352 & 10 Aug 2023 & AU~Mic & 240 / 0.097 & 802.4 & 882 & 0.77 & $V$ & 62$\pm$4 & 2.80$\pm$0.03 & -6.97$\pm$1.21 \\
1166.8834741 & 10 Aug 2023 & AU~Mic & 240 / 0.099 & 802.4 & 850 & 0.59 & $Q$ & & & \\
1166.8938495 & 10 Aug 2023 & AU~Mic & 240 / 0.102 & 802.4 & 772 & 0.65 & $U$ & & & \\
1167.8481084 & 11 Aug 2023 & AU~Mic & 240 / 0.298 & 802.4 & 904 & 0.84 & $V$ & 11$\pm$4 & 2.79$\pm$0.04 & 1.06$\pm$1.25 \\
1167.8590500 & 11 Aug 2023 & AU~Mic & 240 / 0.300 & 802.4 & 907 & 0.63 & $Q$ & & & \\
1167.8697311 & 11 Aug 2023 & AU~Mic & 240 / 0.302 & 802.4 & 905 & 0.64 & $U$ & & & \\
1179.8186449 & 23 Aug 2023 & AU~Mic & 242 / 0.761 & 802.4 & 844 & 0.86 & $V$ & 79$\pm$4 & 2.44$\pm$0.05 & 12.64$\pm$1.16 \\
1179.8302123 & 23 Aug 2023 & AU~Mic & 242 / 0.763 & 802.4 & 860 & 0.63 & $Q$ & & & \\
1179.8420478 & 23 Aug 2023 & AU~Mic & 242 / 0.766 & 802.4 & 880 & 0.66 & $U$ & & & \\
1180.9308369 & 24 Aug 2023 & AU~Mic & 242 / 0.990 & 802.4 & 867 & 0.92 & $V$ & 17$\pm$4 & 2.61$\pm$0.04 & 5.83$\pm$1.13 \\
1181.8186111 & 25 Aug 2023 & AU~Mic & 243 / 0.173 & 802.4 & 866 & 0.84 & $V$ & -16$\pm$4 & 2.88$\pm$0.03 & -9.74$\pm$1.31 \\
1181.8298154 & 25 Aug 2023 & AU~Mic & 243 / 0.175 & 802.4 & 851 & 0.63 & $Q$ & & & \\
1181.8406102 & 25 Aug 2023 & AU~Mic & 243 / 0.177 & 802.4 & 837 & 0.65 & $U$ & & & \\
1182.8234064 & 26 Aug 2023 & AU~Mic & 243 / 0.379 & 802.4 & 830 & 0.94 & $V$ & 27$\pm$5 & 2.50$\pm$0.05 & 15.01$\pm$1.31 \\
1182.8342156 & 26 Aug 2023 & AU~Mic & 243 / 0.382 & 802.4 & 832 & 0.66 & $Q$ & & & \\
1182.8451345 & 26 Aug 2023 & AU~Mic & 243 / 0.384 & 802.4 & 831 & 0.66 & $U$ & & & \\
1184.8133935 & 28 Aug 2023 & AU~Mic & 243 / 0.789 & 802.4 & 894 & 0.90 & $V$ & 71$\pm$4 & 2.40$\pm$0.06 & \\
1185.7993039 & 29 Aug 2023 & AU~Mic & 243 / 0.992 & 802.4 & 902 & 0.82 & $V$ & 14$\pm$4 & 2.57$\pm$0.04 & 6.39$\pm$1.23 \\
1185.8100395 & 29 Aug 2023 & AU~Mic & 243 / 0.994 & 802.4 & 908 & 0.58 & $Q$ & & & \\
1185.8207370 & 29 Aug 2023 & AU~Mic & 243 / 0.996 & 802.4 & 891 & 0.73 & $U$ & & & \\
1186.8327092 & 30 Aug 2023 & AU~Mic & 244 / 0.204 & 802.4 & 740 & 0.91 & $V$ & -20$\pm$4 & 2.83$\pm$0.03 & -10.81$\pm$1.19 \\
1186.8438843 & 30 Aug 2023 & AU~Mic & 244 / 0.207 & 802.4 & 717 & 0.70 & $Q$ & & & \\
1186.8546516 & 30 Aug 2023 & AU~Mic & 244 / 0.209 & 802.4 & 709 & 0.71 & $U$ & & & \\
1187.7936592 & 31 Aug 2023 & AU~Mic & 244 / 0.402 & 802.4 & 670 & 0.95 & $V$ & 16$\pm$5 & & 18.52$\pm$1.16 \\
1187.8247707 & 31 Aug 2023 & AU~Mic & 244 / 0.408 & 802.4 & 691 & 0.71 & $Q$ & & & \\
1187.8355074 & 31 Aug 2023 & AU~Mic & 244 / 0.411 & 802.4 & 650 & 0.74 & $U$ & & & \\
1188.7797490 & 01 Sep 2023 & AU~Mic & 244 / 0.605 & 802.4 & 870 & 0.77 & $V$ & 92$\pm$4 & 2.54$\pm$0.04 & 11.29$\pm$1.07 \\
1188.7903324 & 01 Sep 2023 & AU~Mic & 244 / 0.607 & 802.4 & 882 & 0.64 & $Q$ & & & \\
1188.8013085 & 01 Sep 2023 & AU~Mic & 244 / 0.609 & 802.4 & 896 & 0.61 & $U$ & & & \\
\hline
\end{tabular}} 
\tablefoot{\emr All exposures consist of 4 sub-exposures of equal length.
For each star and each visit, we list the barycentric Julian date BJD, the UT date, the rotation cycle c and phase $\phi$ (computed as indicated in
Sec.~\ref{sec:obs}), the total observing time t$_{\rm exp}$, the peak SNR in the spectrum (in the $H$ band) per 2.3~\kms\ pixel, the noise level in
the LSD Stokes profile, the estimated \Bl\ with error bars (for Stokes $V$ profiles only), the measured small-scale field <$B_s$> and differential
temperature $dT$ with error bars (for each observing night). }
\label{tab:log-au}
\end{table*}

\begin{table*}[ht!]
% \small
\caption[]{Same as Table~\ref{tab:log-au} for our Stokes $VQU$ spectropolarimetric SPIRou observations of EV~Lac in September and October 2023} 
\centering 
\resizebox{\linewidth}{!}{  
\begin{tabular}{ccccccccccc}
\hline
BJD        & UT date & Star & c / $\phi$ & t$_{\rm exp}$ & SNR & $\sigma_P$            & Stokes & \Bl\  &  <$B_s$> &  $dT$ \\
(2459000+) &         &      &            &   (s)        & ($H$) & ($10^{-4} I_c$)      &        &  (G)  &  (kG)    &  (K)  \\
\hline
1213.8123356 & 26 Sep 2023 & EV~Lac & 277 / 0.760 & 1582.4 & 902 & 0.65 & $Q$ & & & \\
1213.8321239 & 26 Sep 2023 & EV~Lac & 277 / 0.765 & 1582.4 & 901 & 0.74 & $U$ & & & \\
1213.8441117 & 26 Sep 2023 & EV~Lac & 277 / 0.768 & 245.2 & 360 & 1.88 & $V$ & 10$\pm$25 & 4.43$\pm$0.07 & 1.81$\pm$0.12 \\
1214.8121319 & 27 Sep 2023 & EV~Lac & 277 / 0.989 & 1582.4 & 813 & 0.82 & $Q$ & & & \\
1214.8321833 & 27 Sep 2023 & EV~Lac & 277 / 0.994 & 1582.4 & 776 & 0.66 & $U$ & & & \\
1214.8443879 & 27 Sep 2023 & EV~Lac & 277 / 0.996 & 245.2 & 291 & 2.24 & $V$ & -105$\pm$30 & 4.39$\pm$0.06 & 2.94$\pm$0.09 \\
1215.8081612 & 28 Sep 2023 & EV~Lac & 278 / 0.217 & 1582.4 & 837 & 0.66 & $Q$ & & & \\
1215.8278955 & 28 Sep 2023 & EV~Lac & 278 / 0.221 & 1582.4 & 837 & 0.66 & $U$ & & & \\
1215.8398404 & 28 Sep 2023 & EV~Lac & 278 / 0.224 & 245.2 & 346 & 1.93 & $V$ & 42$\pm$25 & 4.32$\pm$0.06 & 1.29$\pm$0.09 \\
1216.7910424 & 29 Sep 2023 & EV~Lac & 278 / 0.442 & 1582.4 & 880 & 0.70 & $Q$ & & & \\
1216.8108550 & 29 Sep 2023 & EV~Lac & 278 / 0.446 & 1582.4 & 869 & 0.63 & $U$ & & & \\
1216.8228885 & 29 Sep 2023 & EV~Lac & 278 / 0.449 & 245.2 & 349 & 1.98 & $V$ & -228$\pm$27 & 3.67$\pm$0.06 & 14.58$\pm$0.10 \\
1219.7787074 & 02 Oct 2023 & EV~Lac & 279 / 0.126 & 1582.4 & 698 & 0.75 & $Q$ & & & \\
1219.7986450 & 02 Oct 2023 & EV~Lac & 279 / 0.130 & 1582.4 & 686 & 0.78 & $U$ & & & \\
1219.8103943 & 02 Oct 2023 & EV~Lac & 279 / 0.133 & 245.2 & 266 & 2.53 & $V$ & 132$\pm$37 & 4.34$\pm$0.06 & -1.21$\pm$0.09 \\
1220.7811537 & 03 Oct 2023 & EV~Lac & 279 / 0.355 & 1582.4 & 801 & 0.65 & $Q$ & & & \\
1220.8010876 & 03 Oct 2023 & EV~Lac & 279 / 0.360 & 1582.4 & 752 & 0.71 & $U$ & & & \\
1220.8127874 & 03 Oct 2023 & EV~Lac & 279 / 0.362 & 245.2 & 333 & 1.99 & $V$ & -256$\pm$28 & 3.75$\pm$0.06 & 10.95$\pm$0.09 \\
1221.7787608 & 04 Oct 2023 & EV~Lac & 279 / 0.583 & 1582.4 & 846 & 0.62 & $Q$ & & & \\
1221.7982896 & 04 Oct 2023 & EV~Lac & 279 / 0.588 & 1582.4 & 858 & 0.66 & $U$ & & & \\
1221.8104245 & 04 Oct 2023 & EV~Lac & 279 / 0.590 & 245.2 & 346 & 1.88 & $V$ & 166$\pm$25 & 4.04$\pm$0.06 & 4.93$\pm$0.08 \\
1222.8161946 & 05 Oct 2023 & EV~Lac & 279 / 0.821 & 1582.4 & 891 & 0.70 & $Q$ & & & \\
1222.8363693 & 05 Oct 2023 & EV~Lac & 279 / 0.825 & 1582.4 & 895 & 0.90 & $U$ & & & \\
1222.8485126 & 05 Oct 2023 & EV~Lac & 279 / 0.828 & 245.2 & 358 & 2.00 & $V$ & -33$\pm$30 & 4.25$\pm$0.07 & 0.14$\pm$0.11 \\
1223.8012741 & 06 Oct 2023 & EV~Lac & 280 / 0.046 & 1582.4 & 825 & 0.67 & $Q$ & & & \\
1223.8211277 & 06 Oct 2023 & EV~Lac & 280 / 0.051 & 1582.4 & 786 & 0.78 & $U$ & & & \\
1223.8332969 & 06 Oct 2023 & EV~Lac & 280 / 0.053 & 245.2 & 326 & 2.07 & $V$ & 49$\pm$30 & 4.33$\pm$0.06 & -0.05$\pm$0.09 \\
\hline
\end{tabular}} 
\tablefoot{\emr Longer exposure times were used for Stokes $QU$ observations to ensure detection of the Zeeman signatures. }
\label{tab:log-ev}
\end{table*}

\FloatBarrier %\usepackage{placeins}
\clearpage

\end{appendix}
\end{document}